\documentclass{aa}  

\usepackage{graphicx}
\usepackage{txfonts}
\usepackage{lipsum}
\usepackage{subcaption}         
\usepackage{lscape}             
\usepackage{placeins}           

\usepackage{amsmath}
\usepackage{xcolor}
\usepackage{graphicx}
\usepackage{gensymb}
\usepackage{booktabs}
\usepackage{bm}
\usepackage{lipsum}
\usepackage{algorithm} 
\usepackage{algpseudocode}
\usepackage{multirow}
\usepackage{tabularx} 
\usepackage{array}    
\usepackage[switch]{lineno}

\AtBeginDocument{\mathcode`v=\varv}

\providecommand{\sorthelp}[1]{} 

\newcommand{\HI}{\ifmmode \mathrm{\ion{H}{I}} \else \ion{H}{I} \fi}
\newcommand{\HII}{\ifmmode \mathrm{\ion{H}{II}} \else \ion{H}{II} \fi}
\newcommand{\nh}{\ifmmode N_{{\mathrm{H}} \, \mathrm{I}} \else $N_{{\mathrm{H}} \, \mathrm{I}}$\fi} 
\newcommand{\CODE}{{\tt IViS}}
\newcommand{\BASE}{{\tt Classic3D}}

\newcommand{\norm}[1]{\left\lVert#1\right\rVert}

\newcommand\Ab{\bm{A}}
\newcommand\Bb{\bm{B}}
\newcommand\Gb{\bm{G}}

\newcommand\Ib{\bm{I}}

\newcommand\xb{\bm{x}}

\newcommand\rb{\bm{r}}
\newcommand\Db{\bm{D}}

\newcommand\Sigmab{\bm{\Sigma}}

\newcommand{\Herschel}{\textit{Herschel}}

\def\GHz{\ifmmode $\,GHz$\else \,GHz\fi}
\def\MJysr{\ifmmode \,$MJy\,sr\mo$\else \,MJy\,sr\mo\fi}
\def\microns{\ifmmode \,\mu$m$\else \,$\mu$m\fi}

\def\kms{\ifmmode $\,km\,s$^{-1}\else \,km\,s$^{-1}$\fi}

\usepackage[unicode]{hyperref}
\hypersetup{
  colorlinks=true,      
  linkcolor=blue,       
  citecolor=blue,       
  urlcolor=blue,     
  pdfauthor={Antoine Marchal},
  pdftitle={IViS: Interferometric Visibility-domain Inversion Software}
}

\usepackage[font=small,skip=2pt]{caption}
\usepackage{titlesec}
\titlespacing*{\section}{0pt}{8pt}{4pt}
\titlespacing*{\subsection}{0pt}{6pt}{3pt}

\begin{document}

   \title{IViS: Interferometric Visibility-domain inversion Software}
   \subtitle{A GPU-accelerated Python framework for joint deconvolution with ASKAP.}

    \author{A. Marchal\inst{1,2}
        \and N. M. McClure-Griffiths\inst{2}
        \and J. Dempsey\inst{2}
        \and H. Nguyen\inst{2}
        \and H. Dénes\inst{6}
        \and J. Dickey\inst{3}
        \and C. Lynn\inst{2}
        \and Y. K. Ma\inst{4}
        \and D. McConnell\inst{2} 
        \and M.-A. Miville-Desch\^enes\inst{1}
        \and N. Pingel\inst{5}
        \and J. Th. van Loon\inst{7}
        }

    \institute{Laboratoire de Physique de l’École Normale Supérieure, ENS, Université PSL, CNRS, Sorbonne Université, Université Paris Cité, Observatoire de Paris, F75005, Paris, France \\ 
            \email{antoine.marchal@cnrs.fr}
            \and Research School of Astronomy \& Astrophysics, The Australian National University, Canberra ACT 2611, Australia 
            \and School of Natural Sciences, University of Tasmania, Private Bag 37, Hobart, TAS 7001, Australia 
            \and Max-Planck-Institut f\"ur Radioastronomie, Auf dem H\"ugel 69, 53121 Bonn, Germany 
            \and Indiana University, Department of Astronomy, 727 East Third Street, Bloomington, IN 47405, USA 
            \and College of Sciences and Engineering, Universidad San Francisco de Quito, Quito, Ecuador 
            \and Lennard-Jones Laboratories, School of Chemical and Physical Sciences, Keele University, Keele, Staffordshire ST5 5BG, UK 
            }
   \date{Received June 15, 2026; Revised September 1, 2026; Accepted September 11, 2026.}
 
  \abstract
    {%
    Wide-field spectral-line imaging with modern radio interferometers remains challenging when the emission is extended, multiscale, and distributed over many overlapping pointings. In this regime, accurate reconstruction requires joint treatment of the calibrated visibilities, control of  image-domain regularity, and, when necessary, recovery of missing short spacings from single-dish data.
    We present \CODE\ (Interferometric Visibility-domain Inversion Software), a GPU-accelerated Python framework for visibility-domain joint deconvolution, and assess its performance for wide-field \HI imaging with the Australian Square Kilometre Array Pathfinder (ASKAP).
    The current model, \BASE, reconstructs a non-parametric sky cube directly from calibrated visibilities by minimizing a regularized least-squares criterion. The forward model relies on a non-uniform fast Fourier transform (NUFFT), supports joint deconvolution across all mosaic pointings, and allows positivity constraints through bounded optimization. Short-spacing information can be incorporated through a fusion term that models single-dish data from the same sky estimate. We validate the method using synthetic point-source, noise-only, and multiscale diffuse-emission simulations, and apply it to ASKAP full-survey observations toward the Large Magellanic Cloud (LMC).
    The simulations show that \CODE\ recovers point-source flux accurately in the absence of regularization and that the regularization strength  sets the trade-off between noise suppression and effective resolution. Relative to linear mosaicking of independently deconvolved pointings, joint deconvolution can recover more large-scale power. When single-dish information is included, the reconstructed power spectra closely reproduce the input sky statistics over a broad range of spatial frequencies. Applied to ASKAP data, \CODE\ yields 10\,h mosaics with an effective resolution of 22$^{\prime\prime}$ of Galactic foreground emission and LMC emission with reduced residual side-lobe structure and realistic multiscale morphology.
    A comparison with \texttt{ASKAPSoft} using multi-scale CLEAN shows that, for this specific target and a single observing block, \BASE\ exhibits fewer residual side-lobes and recovers substantially more power at low spatial frequencies.
    \CODE\ provides a flexible visibility-domain framework for joint deconvolution of large spectral-line mosaics, and illustrates the potential of regularized inverse-problem approaches as an alternative to traditional CLEAN-based imaging for next-generation radio surveys.
    }   

   \keywords{methods: data analysis --
          techniques: interferometric --
          techniques: image processing --
          techniques: high angular resolution --
          radio lines: ISM --
          ISM: structure}
          
   \maketitle
   \nolinenumbers
   
\section{Introduction}
Synthesis imaging of large and extended diffuse emission at high angular resolution requires mapping the sky with multiple overlapping pointings, as well as incorporating single-dish data to recover missing short spacings and total power \citep{stanimirovic_2002}. 
In a seminal work on joint deconvolution, \citet{Cornwell:1988} demonstrated that combining visibilities from multiple overlapping pointings \textit{during} the deconvolution process (i.e., deconvolution of a non-linear mosaic of dirty images) recovers more power on spatial scales that are under-sampled by the interferometer, compared to approaches that linearly combine independently deconvolved pointings. 
Using continuum data from the Very Large Array (VLA) at 1390\,MHz toward the \HII region Simeis~57 in the Cygnus~X complex \citep{Gaze:1951,Gaze:1955,Higgs:1991,Oudshoorn:2021}, the author showed that Maximum Entropy methods \citep[hereafter MEM;][]{Cornwell:1985,Narayan:1986} provide a suitable framework for this task. 
By contrast, \citet{Sault:1996} proposed an approach in which deconvolution is performed \textit{after} mosaicking linearly combined dirty images. They showed that this alternative implementation of joint deconvolution yields results comparable to the method proposed by \citet{Cornwell:1988}, while offering advantages in terms of disk space usage because the deconvolution is performed on a single mosaicked image rather than on the full set of individual pointing images.
This work also represented the first application of joint deconvolution to wide-field spectral-line imaging of the 21\,cm line. Imaging of the Small Magellanic Cloud (SMC) at a heliocentric velocity of 123\kms\, required about 20~minutes of computation for both methods using 320 pointings (the total number of visibilities was not reported).
With the method of \citet{Sault:1996} implemented in the \texttt{MIRIAD} package \citep{Sault:1995}, it became a standard approach for deconvolution of interferometric 21\,cm data \citep[e.g.,][]{Staveley-Smith:1997,Kim:1998,Stanimirovic:1999,2005ApJS..158..178M,McClure-Griffiths:2003,McClure-Griffiths:2018,DiTeodoro:2019}, particularly with Australian facilities such as the Australia Telescope Compact Array \citep[ATCA;][]{Frater:1992}.

With the advent of new radio interferometers such as ASKAP and MeerKAT, the volume of calibrated visibilities has increased dramatically. 
While early ASKAP observations with a 16-antenna sub-array remained tractable with \texttt{MIRIAD} \citep{McClure-Griffiths:2018}, the final 36-antenna configuration and longer integration times (up to $\sim$20 hours in its pilot-survey phase) led to a substantial increase in data volume. 
To address these larger datasets, \citet{pingel_2022} used \texttt{WSClean} \citep{offringa-wsclean-2014,offringa-wsclean-2017}, which implements joint deconvolution within a multi-scale CLEAN framework \citep{Cornwell:2008}. Processing a single spectral channel required 4 to 8 hours (on the specific architecture used), depending on source complexity and the number of iterations spent in the major loop, resulting in the imaging of a full ASKAP cube \citep[2048 channels, $\sim$20\,h integration, corresponding to $\sim$3.5\,TB of Stokes~I visibilities;][]{pingel_2022} in about 300 to 600 days of wall-clock processing time distributed on the supercomputer \textit{avatar} at the Australian National University.
Since then, joint deconvolution (with multi-scale CLEAN) has been implemented in \texttt{ASKAPSoft} \citep{Guzman:2019}, and is currently used for the production processing of \textit{full-survey}\footnote{The term \textit{full-survey} observations is used to distinguish these data from those acquired during the \textit{pilot-survey} phase. These early observations consisted of 10-hour integrations over 10 fields that cover the SMC, LMC, and Magellanic Bridge. Full-survey observations covers six fields on the Magellanic Clouds (200 integration hours/field), 20 fields covering a strip along the Galactic Plane (50 hours integration hours/field), 16 fields centered on the Galactic Center (50 integration hours/field), and 48 fields covering a majority of the Magellanic Stream (30 integration hours/field)} GASKAP observations within the ASKAP data reduction pipeline.

Despite these advances, imaging very large spectral-line mosaics remains challenging, particularly when the emission is diffuse, spans a broad range of angular scales, and extends across many overlapping pointings.
While multi-scale CLEAN has significantly improved the recovery of extended emission within the CLEAN framework \citep{Cornwell:2008}, alternative regularized optimization approaches remain attractive because they naturally accommodate prior information, avoid the need for an explicit component-based representation of the sky, and provide a flexible route toward more complex imaging models.
However, applying such methods directly to modern wide-field spectral-line datasets remains computationally demanding.
In this work, we introduce \CODE, a GPU-accelerated imaging software that performs joint deconvolution through visibility-domain optimization, where model visibilities are computed at each iteration via a forward operator. 
The approach revisits regularized visibility-domain imaging formulations in a manner that is scalable to modern interferometric datasets.
Unlike classical MEM implementations, however, our objective is not to reproduce a specific entropy-based reconstruction, but rather to provide a general visibility-domain optimization framework in which different regularization schemes and deconvolution strategies can be readily implemented and compared.
To achieve this goal, we rely on PyTorch \citep{Paszke:2019}, which enables automatic differentiation and efficient GPU-accelerated optimization for large-scale inverse problems. 
It is noteworthy that \CODE\ is not the first successful attempt to provide a flexible deconvolution framework for synthesis imaging. 
In particular, the \texttt{MPol} package \citep{zawadzki_2023} implements a similar forward-modeling approach to predict visibilities at each iteration and was used to image, e.g., ALMA continuum data of protoplanetary disks \citep{Zawadzki:2025}. However, it does not support joint deconvolution of multiple overlapping pointings, which is a key motivation for the present work. 

The paper is organized as follows. 
In Section~\ref{sec:classic3D} we introduce the \BASE\ model, including the forward model and the regularized inverse formulation. 
Section~\ref{sec:workflow} describes the \CODE\ workflow and software architecture. 
In Section~\ref{sec:validation} we validate the method using synthetic observations of increasing complexity. 
Section~\ref{sec:application} presents its application to GASKAP-HI\footnote{GASKAP-HI is a large program of 21\,cm observations of the Milky Way (hereafter MW), and the Magellanic system with ASKAP.} (PIs: Naomi McClure-Griffiths, John Dickey, Nickolas Pingel) data. 
Finally, Section~\ref{sec:conclusions} summarizes the main findings and Section~\ref{sec:prospects} outlines future prospects.

\section{Classic3D} \label{sec:classic3D}
Our aim is to model the data measured by a radio interferometer, namely the spatial
coherence function of the electric field
\begin{align}
    \mathcal{V}_{i, j}
    = \big\langle E\big(\xb_i, t\big) \, E^*\big(\xb_j, t\big)\big\rangle_t \, ,
\end{align}
where $\langle \cdot \rangle_t$ denotes the time average, $\xb_i$ and $\xb_j$ are the
positions of antennas $i$ and $j$ in the array reference frame, and $\mathcal{V}_{i,j}$
is the visibility measured by the baseline formed by these two antennas.

For that, we have implemented, as a base layer, the model class \BASE, where the \texttt{3D} name refers to the ability of \CODE\, to natively support the spectral dimension of the data when present. 
In other words, the model presented below is built over a 3D sky model with two spatial dimensions and one spectral dimension.
While this general formalism is not needed when each frequency channel is treated independently (which is usally the case for spectral line data), it becomes necessary when fitting multiple frequencies at once, e.g., for multi-frequency synthesis \citep[MFS;][]{Conway:1990,Sault:1994,Rau:2011}.
Importantly, as described below, \BASE\, demonstrates the use of spatial regularization using a Laplacian filtering to obtain a spatially coherent solution of the image cube of the sky.

\subsection{Forward model} \label{subsec:forward}
\begin{figure}
    \includegraphics[width=\linewidth]{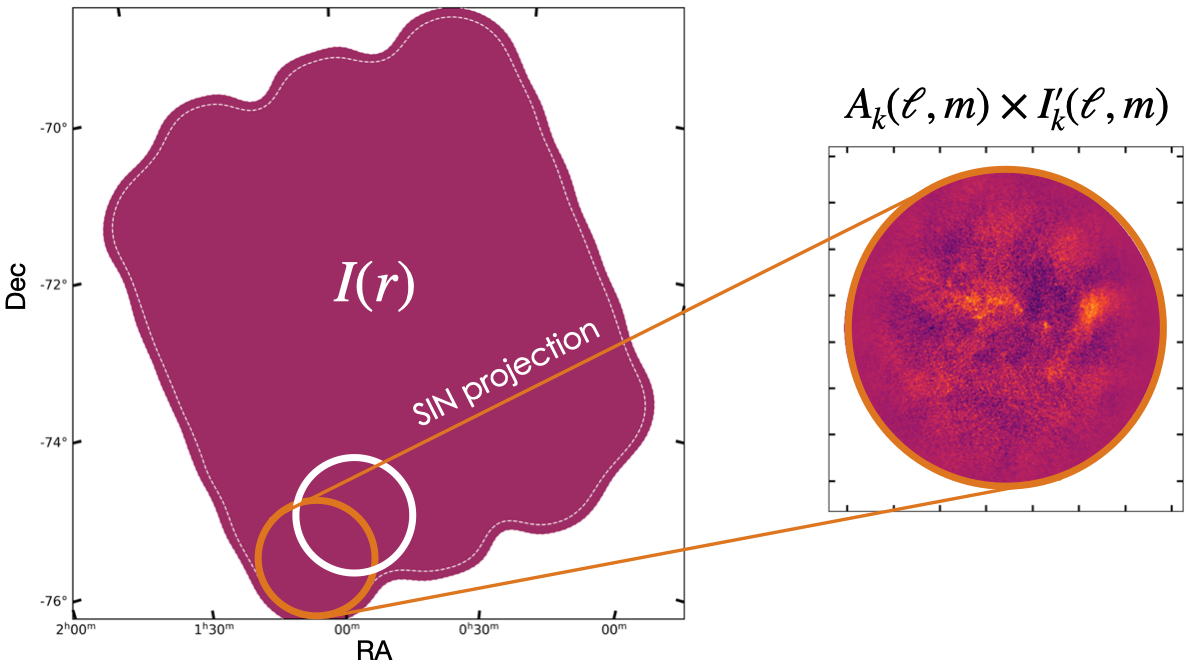}
  \caption{
    Illustration of the SIN projection at the position of pointing $k$ (orange circle), transforming the image model $\Ib_{\nu}(\rb)$, defined on the arbitrary WCS projection chosen to image the data, into a local sub-image $\Ib'_{k,\nu}(\ell,m)$.
    The left panel shows the initial image model before optimization (here a null prior). 
    The inset illustrates the sky brightness distribution sampled by pointing $k$ in its local coordinate system (arbitrary units).
    The white circle indicates the overlap between adjacent mosaic pointings.}
  \label{fig:SIN_proj}
\end{figure}
For an image cube of the sky $\Ib_{\nu}(\rb)$ at frequency $\nu$ and sky coordinates $\rb$, we model a set of calibrated visibilities for a pointing $k$ as 
\begin{align} \label{eq:vis}
    \tilde{\bm{\mathcal{V}}}_{k,\nu}(\Ib_{k,\nu}'; u, v, w) = \iint &\frac{\Ab_{k,\nu}(\ell,m)\Ib_{k,\nu}'(\ell, m)}{\sqrt{1-\ell^2-m^2}} \notag\\ & \times \, e^{-2 \pi i\left[u \ell+v m+w\left(\sqrt{1-\ell^2-m^2}-1\right)\right]} d \ell d m \, ,
\end{align}
where $\Ab_{k,\nu}(\ell,m)$ is the primary beam that depends on the frequency $\nu$, $\ell$ and $m$ are direction cosines \citep{white_book_1999}, and %
\begin{align} \label{eq:sin_projection}
\Ib_{k,\nu}'(\ell,m) \xleftarrow{\textbf{SIN projection}} \Ib_{\nu}(\rb)
\end{align}
is a sub-region of $\Ib_{\nu}(\rb)$, an orthographic/synthesis projection \citep{calabretta_2002} towards the pointing center of each pointing $k$, and whose extent is that of the primary beam of the instrument. 
Note that $\Ib_{\nu}(\rb)$ is defined in an arbitrary WCS projection (not necessarily orthographic) and so a (re-)projection from the parent coordinate system to an orthographic projection is performed. 
Figure~\ref{fig:SIN_proj} illustrates the SIN (re-)projection for one pointing of an ASKAP mosaic.
In practice, $N$ grids are pre-computed to facilitate these (re-)projections performed with bilinear interpolations during the optimization process, with $N$ the total number of observed pointings (e.g., 108 for the three-interleave ASKAP footprint used in GASKAP-HI, corresponding to 36 pointings per interleave).

In the small angle approximation, and/or coplanar array, Equation~\ref{eq:vis} simplifies to
\begin{align}
    \tilde{\bm{\mathcal{V}}}_{k,\nu}(\Ib_{k,\nu}'; u, v)
    \simeq
    \iint
    \Ab_{k,\nu}(\ell,m)\,
    \Ib_{k,\nu}'(\ell, m)\,
    e^{-2 \pi i (u \ell + v m)}
    \, d\ell\, dm \, ,
\end{align}
In this simplified framework, the projected sky models modulated by the primary beams $\Ab_{k,\nu}(\ell,m) \times \Ib_k'(\ell,m)$ are directly transformed into the set of visibilities $\bm{\mathcal{V}}_{k,\nu}$ via Fourier transforms, neglecting the w-term.
This approximation is partly mitigated by modeling each pointing locally around its own pointing center over an angular extent limited by the primary beam, thereby reducing the impact of non-coplanar baseline effects compared to a single global projection of the full mosaic.
The associated phase error nevertheless scales as $\pi |w|\theta^2$, where $\theta$ is the angular distance from the phase center.
In practice, for the diffuse \HI emission and angular scales considered in this work, this local tangent-plane treatment is expected to provide an adequate approximation and is consistent with the classical framework of interferometric mosaicing \citep[e.g.,][]{Sault:1996,Staveley-Smith:1997,Kim:1998,Stanimirovic:1999,2005ApJS..158..178M,McClure-Griffiths:2003,McClure-Griffiths:2018}.

Following \citet[][{\tt MPoL} package\footnote{\href{https://mpol-dev.github.io/MPoL/}{https://mpol-dev.github.io/MPoL/}}]{zawadzki_2023}, we use a non-uniform fast Fourier transform to perform this transformation. 
While {\tt MPoL} leveraged {\tt torchkbnufft\footnote{\href{https://torchkbnufft.readthedocs.io/}{https://torchkbnufft.readthedocs.io/}}}, we instead chose the Flatiron Institute's implementation, {\tt fiNUFFT\footnote{\href{https://finufft.readthedocs.io/}{https://finufft.readthedocs.io/}}} \citep{Barnett:2019,Barnett:2020}. 
Specifically, we made use of the PyTorch wrapper {\tt Pytorch-FINUFFT\footnote{\href{https://flatironinstitute.github.io/pytorch-finufft/}{https://flatironinstitute.github.io/pytorch-finufft/}}}, allowing us to make use of the auto-differentiation available in PyTorch to calculate the gradient of the cost function described below, while also enabling the use of {\tt cuFINUFFT} \citep{Shih:2021} for GPU acceleration.
The choice of a NUFFT to model the visibilities, as opposed to a uniform FFT, is also motivated by our desire to eliminate the need to explicitly implement a de-gridding step during the optimization process.
This operation is handled internally by the NUFFT and is fully self-contained within the transforms provided by {\tt cuFINUFFT}.

\subsection{Cost function}\label{subsec:cost}
The residual visibilities for each pointing $k$ is
\begin{align}
    \bm{L}_{k,\nu}(\Ib_{k,\nu}') = \tilde{\bm{\mathcal{V}}}_k(\Ib_{k,\nu}')  - \bm{\mathcal{V}}_{k,\nu} \,
\end{align}
and the estimated parameter maps $\Ib_{k,\nu}(\rb)$ are defined as the minimizer of a cost function that includes the sum of the squares of the residual
\begin{align}
  J_{k}(\Ib_{k,\nu}') = \frac{1}{2} \, \norm{L_{k,\nu}\big(\Ib_{k,\nu}'\big)}_{\bm{\Sigmab_{k,\nu}}}^2 = \frac{1}{2} \, 
  \sum_{\nu,u,v} \left(\frac{L_{k,\nu}\big(\Ib_{k,\nu}'\big)}{\bm{\Sigmab_{k,\nu}}}\right)^2 \, ,
\end{align}
where $\Sigmab_{k,\nu}$ is the standard deviation of the noise for each measured visibility, provided in the measurement set of each pointing $k$ by the data column {\tt SIGMA};
summed over $k=1$ to $N$ 
\begin{align} \label{eq:Q}
    Q(\Ib) = \sum_{k}^{N} J_k(\Ib_k') \, .
\end{align}
It is this sum over all overlapping pointings contained in the sky coverage of $\Ib_{k}$ that makes this deconvolution \textit{joint}.

To ensure spatial coherency of the sky images $\Ib_{\nu}(\rb)$ at the scale of the grid pixelation, we introduce a Laplacian filtering of $\Ib_{\nu}(\rb)$ as a regularizer\footnote{Note that regularizers lie at the core of regularized maximum-likelihood methods, and that there exists a whole family of such regularizers, described comprehensively in \citet[][see their section~3.1]{zawadzki_2023}.} to penalize its high spatial frequencies.
The kernel used is
\begin{equation}
    d = \begin{bmatrix}
        0 & -1 & 0 \\
        -1 & 4 & -1 \\
        0 & -1 & 0 \\
        \end{bmatrix} \, ,
\end{equation}
and the regularization term is 
\begin{equation}
  \label{eq:3}
  R(\Ib) = \sum_{\nu} \frac{1}{2} \, \|\Db \Ib_{\nu}(\rb)\|_2^2 \, ,
\end{equation}
where $\Db$ is the matrix that performs the convolution with the kernel $d$.
The total cost function is  
\begin{align} \label{eq:total_cost}
    Q_{\rm tot}(\Ib) = Q(\Ib) + \lambda_r\, R(\Ib) \,.
\end{align}
where we have introduced a hyper-parameter $\lambda_r$ that tunes the balance between the two terms. 
It is instructive already to consider the role that $R(\Ib)$ plays in shaping the effective instrumental response of the reconstructed images.
For low values of $\lambda_r$, the high–spatial-frequency content of the reconstruction will always likely be dominated by noise present in the data, whereas for large values of $\lambda_r$, the Laplacian regularization leads to overly smooth parameter maps, effectively favoring smoothness over data fidelity.
As is clear from Equation~\ref{eq:total_cost}, the regularization term $R(\Ib)$ and its associated hyperparameter $\lambda_r$ therefore play a central role in determining the instrumental response of the reconstructed (i.e., deconvolved) sky images.
By analogy with a traditional {\tt CLEAN} approach, one might be tempted to refer to this as the \CODE\, \textit{restoring} beam. However, there is a fundamental difference: \CODE\, does not include an explicit convolution step in which the size of the \textit{restoring} beam is prescribed by the user.
This will be discussed further in Section~\ref{subsec:noise} where the beam \textit{response} or \textit{effective beam} (as opposed to a \textit{restoring} beam) of \CODE\, will be quantified using a power spectrum analysis.

Finally, the total cost function is a regularized non-linear least-square criterion and the minimizer is
\begin{equation}
    \hat \Ib(\rb) = \underset{\Ib}{\text{argmin}}\ Q_{\rm tot}(\Ib) \, .
\end{equation}
We propose a solution that relies on an iterative optimization algorithm (see Section~\ref{subsec:optimization}) that uses the gradient of the cost function $\nabla Q_{\rm tot}\left(\Ib\right)$.

\subsection{Conceptual comparison with \texttt{MEM}} \label{subsec:comparison_MEM}
It is useful at this stage to compare, at least conceptually, \BASE\ with the \texttt{MEM}, which has been widely used to image diffuse emission lines within a joint deconvolution framework \citep{Sault:1996}, including with 21\,cm data \citep[e.g.,][]{Staveley-Smith:1997,Kim:1998,Stanimirovic:1999,McClure-Griffiths:2005} that is of interest here.
\BASE\ and \texttt{MEM} share the same data-fidelity term (i.e., Equation~\ref{eq:Q}); however, depending on the specific implementation, this term may be evaluated either directly from all visibilities or from linearly mosaicked, gridded images for computational practicality \citep{Sault:1996}.
Beyond the data-fidelity term, the Laplacian regularization used here and the maximum-entropy\footnote{A comprehensive description of the maximum entropy regularizer can be found in \citet[][see their section 3.1.2]{zawadzki_2023}} term employed in \texttt{MEM}, while functionally different, serve a similar purpose: enforcing spatial coherence by preventing high spatial frequencies from being unrealistically driven by noise. 
In interferometric imaging, these small-scale modes are generally less constrained by the data because of the increasingly sparse sampling of the $uv$-plane at long baselines.
In both cases, the solution is encouraged to remain spatially smooth and coherent on scales of a few pixels.

A key difference, however, lies in how positivity is enforced. In \texttt{MEM}, the entropy term simultaneously promotes smoothness and favors positivity, whereas in \BASE\ these two aspects are decoupled: spatial smoothness is imposed through the Laplacian regularization, while positivity is enforced explicitly through bounded optimization (see Section~\ref{subsec:optimization}).
This decoupling between smoothness and positivity is an important advantage, as it potentially allows the framework to be extended in the future to non–strictly positive Stokes parameters (e.g., Q, U, and V).
Finally, in \texttt{MEM} as in \BASE, and as discussed above, these regularizers ultimately set the effective beam response of the reconstructed image $\hat{\Ib}(\rb)$.

\subsection{Short spacing: fusion and feathering}
\label{subsec:fusion}
Synthesis imaging of resolved emission requires the addition of visibilities at the zeroth spatial frequency (total power) obtained with a single-dish telescope with overlapping low spatial frequency. Due to sparse $uv$-coverage from, e.g., limited elevation range or observing duration, interferometer visibilities at these spatial frequencies are either entirely missing or poorly constrained.
These large-scale components can only be recovered through single-dish total-power measurements, or, when available, the auto-correlations of individual antennas.
Traditional \textit{feathering} solves this problem by performing a post-processing combination of independently reconstructed single-dish and interferometric images.
In contrast, we adopt here the term \textit{fusion} to denote the incorporation of single-dish information directly within the deconvolution process.
For instance, \citet{stanimirovic_2002} used an approach that combines the dirty image with single-dish data (and merged beams) prior to deconvolution, which classifies as fusion.

In \BASE, we implemented the following fusion strategy:
data observed at low resolution with a single-dish telescope are modeled from the same sky image model $\Ib_{\nu}(\rb)$.
The brightness temperature is 
\begin{align}
    \tilde{T}_{b,\nu}(\rb) = \Bb_{\nu}(\rb) \otimes \Ib_{\nu}(\rb) \, ,
\end{align}
where $\Bb_{\nu}(\rb)$ is the beam of the single dish telescope, here assumed to be Gaussian, and $\otimes$ denotes the convolution operation.
While an idealized Gaussian beam is currently used for the single-dish component in \CODE, image-based beam models could readily be incorporated, as already done for the interferometric primary beams, allowing realistic beam asymmetries and side-lobe structures to be included if needed.
The residual brightness temperature is
\begin{align}
    L_{\rm sd,\nu}(\Ib) = \tilde{T}_{b,\nu}(\Ib)  - T_{b,\nu} \, ,
\end{align}
and like before $\Ib(\rb)$ is defined as the minimizer of a cost function that is the sum of $Q(\Ib)$ and 
\begin{align}
  K(\Ib) = \frac{1}{2} \, \norm{L_{\rm sd}\big(\Ib\big)}_{\bm{\Sigmab_{\rm sd,\nu}}}^2 \, ,
\end{align}
where $\bm{\Sigmab_{\rm sd}}$ is the standard deviation of the noise in the single-dish data, typically estimated from emission-free channels in spectral-line\footnote{In single-dish continuum data, such a data-driven noise estimate may be more difficult because signal-free regions are not always available.} observations.
The total cost function becomes
\begin{align} \label{eq:fusion_cost}
    Q_{\rm tot}(\Ib) = Q(\Ib) + \lambda_s \, K(\Ib) + \lambda_r\, R(\Ib) \,,
\end{align}
where $\lambda_s$ is a hyper-parameter that tunes the balance with the other terms (i.e., data fidelity and regularization).
In contrast to traditional feathering approaches, the formalism proposed here models the data from each instrument independently while using a common sky model $\Ib_{\nu}(\rb)$.
The \textit{fusion} therefore occurs directly during the optimization process through the joint minimization of $Q(\Ib)$ and $K(\Ib)$.
Conceptually, this approach is related to previous efforts to incorporate single-dish information directly within the deconvolution process, including MEM-based methods \citep{stanimirovic_2002}, the GILDAS approach in which single-dish data are converted into pseudo-visibilities and merged with the interferometric visibilities prior to imaging and deconvolution \citep{RodriguezFernandez:2008}, and the SDINT framework of \citet{Rau:2019}.
The implementation proposed here differs in that the interferometric and single-dish data enter explicitly as separate terms in a common visibility-domain optimization problem, rather than being combined through iterative image-domain major/minor deconvolution cycles.
A caveat of this approach, which also applies to traditional feathering methods, is that the spectral resolution of the two datasets should ideally be matched. In practice, however, this is not always the case. For example, GASS (the Parkes\footnote{Hereafter, we use Parkes to refer to CSIRO's 64-m radio telescope, now officially named Murriyang.} Galactic All-Sky Survey) data have a spectral resolution of 0.82\,km\,s$^{-1}$ \citep{McClure-Griffiths:2009,Kalberla:2010,Kalberla:2015}, whereas ASKAP calibrated visibilities are typically delivered with a channel spacing of 0.5\,km\,s$^{-1}$.
In such situations, it is commonly assumed that the short-spacing modes vary smoothly with frequency and can therefore be used to reconstruct interferometric image cubes at higher spectral resolution.
A possible alternative would be to explicitly model the spectral response of each instrument within the forward operators and cost functions entering Equation~\ref{eq:fusion_cost}. This is beyond the scope of the present work, and in the ASKAP application presented below we adopt the assumption that the short-spacing modes vary smoothly with frequency.
Finally, although short-spacing information can be incorporated into \CODE, it is not required: setting $\lambda_s=0$ simply ignores the provided single-dish data.

\section{IViS Workflow} \label{sec:workflow}
\begin{figure}
    \includegraphics[width=\linewidth]{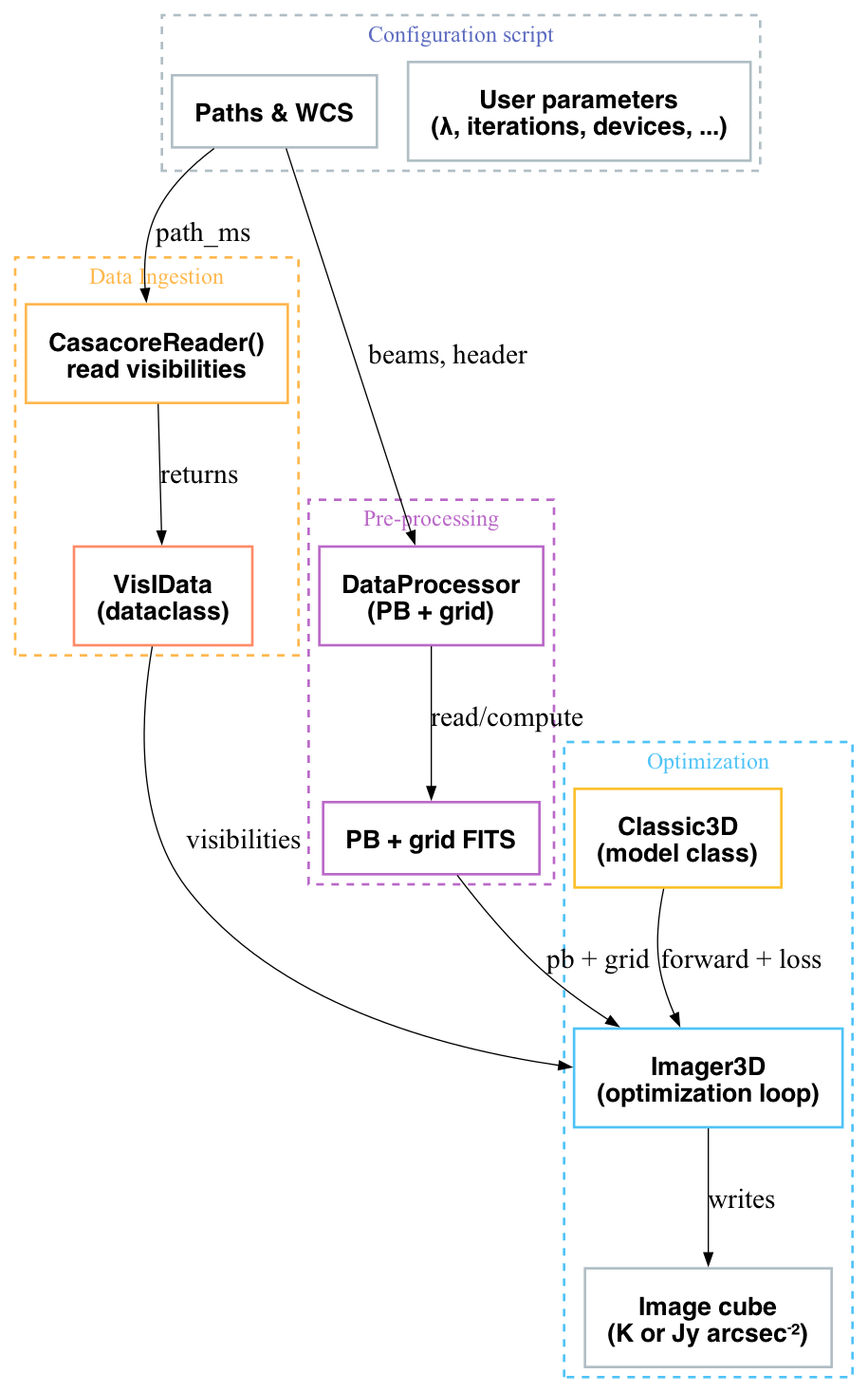}
  \caption{
    Modular imaging workflow of \CODE. 
    Calibrated visibilities are read with the \texttt{CasacoreReader} which returns a \texttt{VisIData} dataclass object, primary beams and interpolation grids used for the joint deconvolution are provided by the \texttt{DataProcessor}, and all inputs are passed to the \texttt{Imager3D} together with a chosen model class (e.g., \BASE). 
    The output is an optimized image cube in physical units of K or Jy\,arcsec$^{-2}$.
    }
  \label{fig:workflow}
\end{figure}
\begin{table*}
\centering
\small
\caption{Summary of the simulations and datasets used in this work.}
\label{tab:experiments_summary}
\begin{tabularx}{\textwidth}{
p{0.12\textwidth}@{\hspace{2pt}}
X@{\hspace{2pt}}
X@{\hspace{4pt}}
X@{\hspace{4pt}}
X}
\toprule
Section & Purpose & Sky model/data & Interferometric setup & Short spacings \\
\midrule
Sect.~\ref{subsec:point-source} & Flux validation & Point source & VLA D & None \\
Sect.~\ref{subsec:noise} & Noise/effective beam & Noise-only visibilities & MeerKAT-like & None \\
Sect.~\ref{subsec:synthetic} & Joint deconvolution & fBm image & ASKAP footprint / MeerKAT layout & Parkes-like simulation \\
Sect.~\ref{subsec:synthetic} & Realistic mock sky & Herschel 250\,$\mu$m map & ASKAP footprint / MeerKAT layout & Parkes-like simulation \\
Sect.~\ref{subsec:results} & Real-data application & GASKAP-HI LMC fields & ASKAP & GASS/Parkes \\
Sect.~\ref{subsec:askapsoft} & Comparison with \texttt{ASKAPSoft} & GASKAP-HI LMC fields & ASKAP & GASS/Parkes \\
Sect.~\ref{subsec:atca} & External comparison & LMC H\,I cube & ATCA & Parkes \\
\bottomrule
\end{tabularx}
\end{table*}

\CODE\, was designed to support the flexible implementation of multiple models within a single deconvolution framework, with the \BASE\, model presented above acting as a foundational layer upon which additional complexity can be introduced.
This modeling approach is embedded in a larger, modular imaging workflow in which the user explicitly assembles the main components of the reconstruction, as illustrated in Figure~\ref{fig:workflow}.
These components are described as follows. 

\subsection{Configuration script}
The primary user-facing entry point to \CODE\, is a single parent configuration script, illustrated by the first box (top) in Figure~\ref{fig:workflow}. This script allows the user to control the main parameters of the run (e.g., hyper-parameters, number of optimizer iterations, and computing device such as CPU or GPU), specify the paths to the raw data (i.e., the visibilities) and primary beam images $\Ab_{k,\nu}(\ell,m)$, and define the WCS of the sky image used to perform the deconvolution.
A notebook illustrating a basic usage example is available at \url{https://ivis-dev.readthedocs.io/en/latest/tutorials/get_started.html}.

\subsection{Data ingestion}
From this configuration script, data ingestion is handled by a set of \texttt{Reader} classes, implemented in the \texttt{ivis.readers} module. 
Note that Inyarrimanha Ilgari Bundara, the CSIRO Murchison Radio-astronomy Observatory currently use \texttt{Common Astronomy Software Applications} \citep[hereafter \texttt{CASA};][]{CASATeam:2022} \texttt{MeasurementSets} \citep[v2,][]{kemball2000ms} as the standard format to store calibrated visibilities and associated metadata for ASKAP; however, this standard is likely to evolve over the coming decades.
Alternative formats, such as \textit{Zarr} combined with \textit{xarray} data structures \citep{hoyer2017xarray}, may become more widely adopted in the future.
For this reason, \CODE\, has been designed so that these \texttt{Reader} classes can be easily replaced by new implementations adapted to emerging data formats, provided they return the internal \texttt{VisIData} dataclass defined in \CODE.
At present, visibilities are ingested using the \texttt{CasacoreReader} class, which relies on the \texttt{casacore} library to produce structured \texttt{VisIData} objects. This is illustrated with the yellow box in Figure~\ref{fig:workflow}.
As the name suggests, the current implementation supports Stokes~I only.

The ingestion framework was also designed to facilitate both large mosaic imaging and the combination of multiple observing rounds.
Measurement Sets associated with different pointings and independent observing blocks are automatically discovered within a parent directory and concatenated internally into a single \texttt{VisIData} object prior to imaging.
This enables the reconstruction of a single sky model constrained simultaneously by all available visibilities, allowing deeper integrations to be formed from multiple observing rounds.
A warning is issued if inconsistencies in the phase centers of the Measurement Sets are detected.
The ASKAP application presented in this paper combines multiple adjacent fields from a single observing round only; the combination of repeated observing rounds for deeper integrations, although straightforward within the present framework, is left for future work.

\subsection{Pre-processing}
Before the optimization can be performed, two pre-processing steps are required. 
\CODE\ does not assume that the primary-beam images $A_{k,\nu}(\ell,m)$ provided by the user share the same reference frame or angular pixel scale as the reconstructed sky image $I_{\nu}(\rb)$.  
We therefore provide a \texttt{DataProcessor} class that uses the WCS information of both datasets to make them compatible. 
In practice, each primary beam is first resampled to the pixel scale of the target image and, when necessary, reprojected onto the corresponding beam-plane WCS. 
A second product is then precomputed for the forward operator: an interpolation grid that specifies, for each pixel of the primary-beam image, the corresponding sampling location in the sky-image frame. 
During the optimization, this grid is used together with the bilinear interpolation described in Equation~\ref{eq:sin_projection} to evaluate the current sky estimate consistently in the primary-beam frame before multiplication by the primary beam. 
Because these operations can be time-consuming, \texttt{DataProcessor} can write the reprojected beams and interpolation grids to disk in FITS format for reuse across repeated executions of the configuration script. 
These steps are illustrated by the purple box in Figure~\ref{fig:workflow}.

\subsection{Optimization} \label{subsec:optimization}
Finally, all inputs are passed to the \texttt{Imager3D}, which centralizes the optimization. 
The imager receives the visibility data, primary beams and interpolation grids for all pointings, together with the image properties defined by the target WCS. 
A model such as \BASE, provides the forward operator mapping images to visibilities and the corresponding loss function that will be used by the optimizer.
The result is an image cube in physical units of K or Jy\,arcsec$^{-2}$. 
This is illustrated in Figure~\ref{fig:workflow} with the cyan box.

The optimization implemented in \texttt{Imager3D} relies on L-BFGS-B \citep[Limited-memory Broyden-Fletcher-Goldfarb-Shanno with Bounds;][]{zhu_algorithm_1997}, a quasi-Newton iterative algorithm. 
In LBFGS-B, after an initialization $I_{(0)}$, the
solution is approached iteratively with a gradient descent
\begin{equation} \label{eq:L-BFGS-B}
    \Ib_{(k+1)} = \Ib_{(k)} - \alpha_{(k)} \bm{H}^{-1}_{(k)} \nabla Q_{\rm tot}\left(\Ib_{(k)}\right) \, ,
\end{equation}
where $\bm{H}^{-1}$ is approximated with the L-BFGS formula, and $\alpha$ is the step that is internally determined by the algorithm; until reaching the maximum number of iteration set by the user. 
It is important to note that $\nabla Q_{\rm tot}(\Ib)$ is not analytically derived here. 
This could be done for further optimization of the computing time but instead we have relied on the built-in auto-differentiation of the PyTorch framework to allow for more flexible algorithmic development of new models. 
Note as well that the L-BFGS-B is called from {\tt scipy.optimize} and is a wrapper around a {\tt Fortran 77/90} code.

LBFGS-B was chosen not only for its ability to handle large-scale optimization problems, but also, as its name suggests, for its capacity to impose explicit bounds on the parameter space during the search.
As mentioned in Section~\ref{subsec:comparison_MEM}, being able to bound the parameter space is critical for synthesis imaging: while Stokes~Q, U, and V are expected to take both positive and negative values, the \HI\ emission component of Stokes~I is generally expected to remain positive.
Continuum-subtracted \HI\ data may nevertheless contain negative values in the presence of absorption features.
Imposing this prior knowledge on $\Ib_{\nu}(\rb)$ naturally leads to improved deconvolution performance.
For the applications considered in this paper, namely spectral-line imaging in Stokes~I, a hard positivity constraint is therefore enforced by bounding the search space such that $\Ib_{\nu}(\rb) \ge 0$.
This positivity constraint should be understood as a prior on the emission component of the sky brightness; absorption features are not explicitly modeled by this positivity-constrained version of \BASE.

\section{Validation} \label{sec:validation}
Before applying \CODE\, to calibrated data from ASKAP, we designed a three-level validation procedure.
First, we verify that all unit conversions in \CODE\, are correctly implemented and that the flux of a single point source can be accurately recovered in an idealized setup, using a synthetic observation generated with \texttt{CASA}.
Second, we provide a comprehensive characterization of \CODE’s effective beam response as well as its noise properties.
Finally, we assess the joint deconvolution performance of \CODE\, using synthetic observations of multiple pointings derived from a known single-frequency image $\Ib_{\nu}(\rb)$, and evaluate its ability to recover the multi-scale structure of the input signal via a power-spectrum analysis.
All synthetic measurement sets used in this paper were generated using \texttt{CASA}. 
Table~\ref{tab:experiments_summary} summarizes the different simulations, interferometric configurations, and observational datasets used throughout the validation and application sections.

\subsection{Point-source simulation}
\label{subsec:point-source}
We made use of the NRAO-provided ``Simulation in CASA'' community examples\footnote{Available at \url{https://casadocs.readthedocs.io/en/stable/examples/community/simulation_script_demo.html}}
 to simulate observations of the sky over multiple frequencies.
We simulated a single-pointing interferometric observation using the \texttt{CASA} simulator.
The antenna configuration corresponds to the VLA D configuration, with ideal alt-az mounted antennas and perfect circular polarization feeds.
A single spectral window was defined in L band, centered at 1.0\,GHz with a bandwidth of 1.0\,GHz, sampled with 10 frequency channels in dual circular polarization (RR, LL).
The observation was phased to the arbitrarily chosen J2000 position $19^{\mathrm h}59^{\mathrm m}28.5^{\mathrm s}$, $+40^{\circ}44^{\prime}01.5^{\prime\prime}$.
Auto-correlations were excluded, and standard shadowing and elevation limits were applied.
Visibilities were generated over a $\pm5$\,h hour-angle range around transit with an integration time of 2000\,s.
The sky model consisted of a single component located at the phase center
(J2000 $19^{\mathrm h}59^{\mathrm m}28.5^{\mathrm s}$, $+40^{\circ}44^{\prime}01.5^{\prime\prime}$), and defined as a point-source.
The total flux density was set to 5\,Jy at a reference frequency of 1.5\,GHz, assuming a power-law spectrum with spectral index $\alpha=-1.0$, where $S_\nu \propto \nu^\alpha$.
An additional Gaussian random noise with a standard deviation of 0.1\,Jy was added independently to each complex visibility sample.
\begin{figure}
    \includegraphics[width=\linewidth]{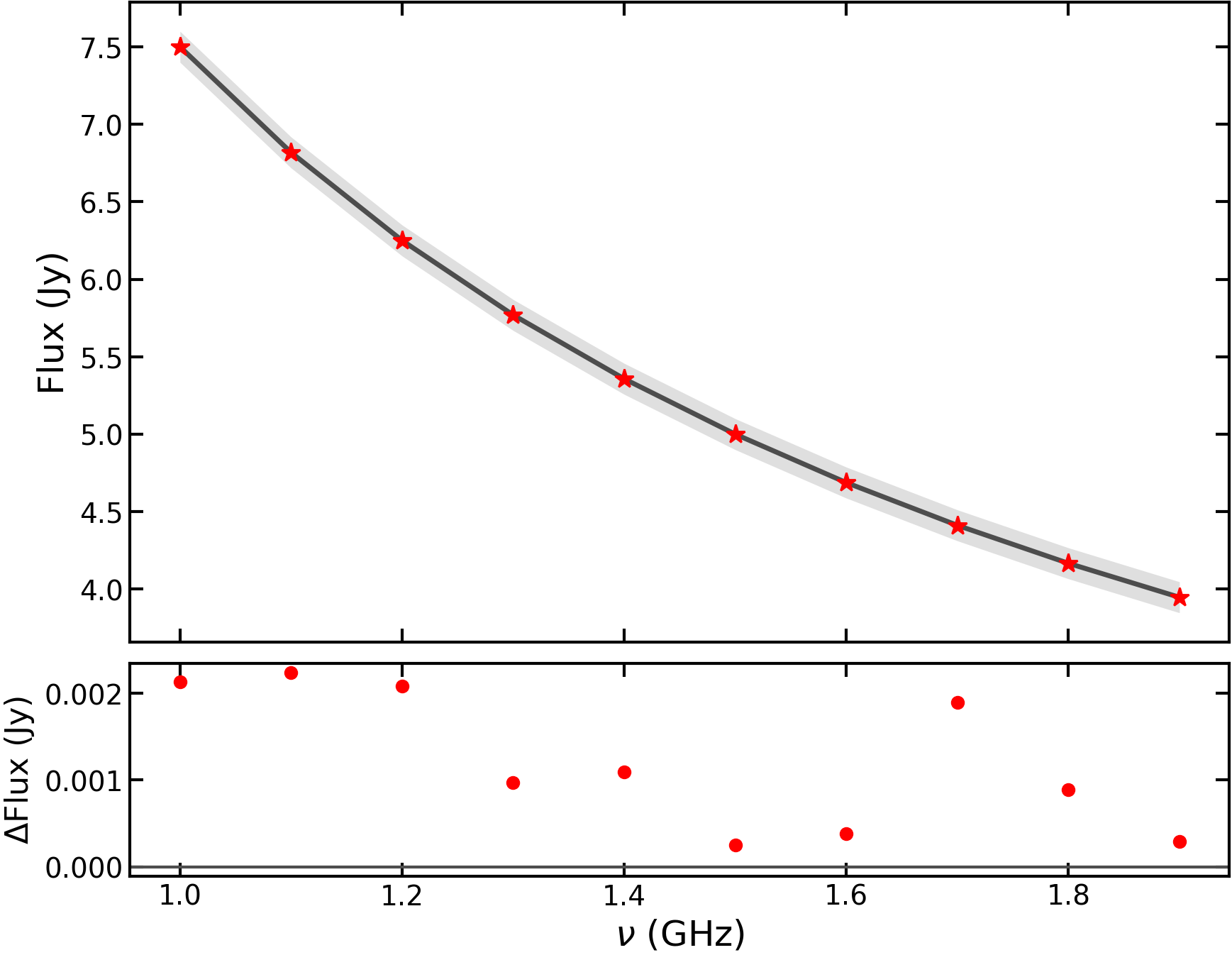}
  \caption{
    Flux versus frequency for a single point-source synthetic observation made with \texttt{CASA}.
    The black line shows the input flux.
    The gray shaded area illustrates the amplitude of the Gaussian noise injected independently into each complex visibility sample.
    The red stars show the recovered flux from \CODE\, (\BASE) with $\lambda_r=0$.
    The lower panel shows the flux residual (true - recovered).
    }
  \label{fig:point_source_flux_residual}
\end{figure}

Figure~\ref{fig:point_source_flux_residual} shows a scatter plot of the flux vs frequency for the point-source simulation. 
The black line shows the input flux, and the gray shaded area illustrates the $\pm0.1$\,Jy Gaussian random noise added to the visibilities.
The red stars show the recovered flux from \CODE\, (\BASE) with $\lambda_r=\lambda_s=0$\footnote{Note that no single dish data are needed to measure the flux of a point source, and so we set $\lambda_s=0$.}, and $\Ib_\nu(\rb)>0$.
The lower panel shows the flux residual (true - recovered).
This choice of $\lambda_r$ is motivated by the fact that for a point-source, all the flux should be accumulated in a single pixel of $\Ib_\nu(\rb)$. 
Introducing a regularizer inevitably leads to a redistribution of flux over adjacent pixels, with the characteristic scale of this spreading being a non-trivial function of the regularization strength $\lambda_r$. Importantly, this redistribution does not, in general, preserve total flux, and can therefore bias the recovered photometry of unresolved sources. For this reason, we caution users against adopting a non-zero value of $\lambda_r$ when accurate point-source flux recovery is required.

\subsection{Noise-only synthetic observation} \label{subsec:noise}
\begin{figure}
    \includegraphics[width=\linewidth]{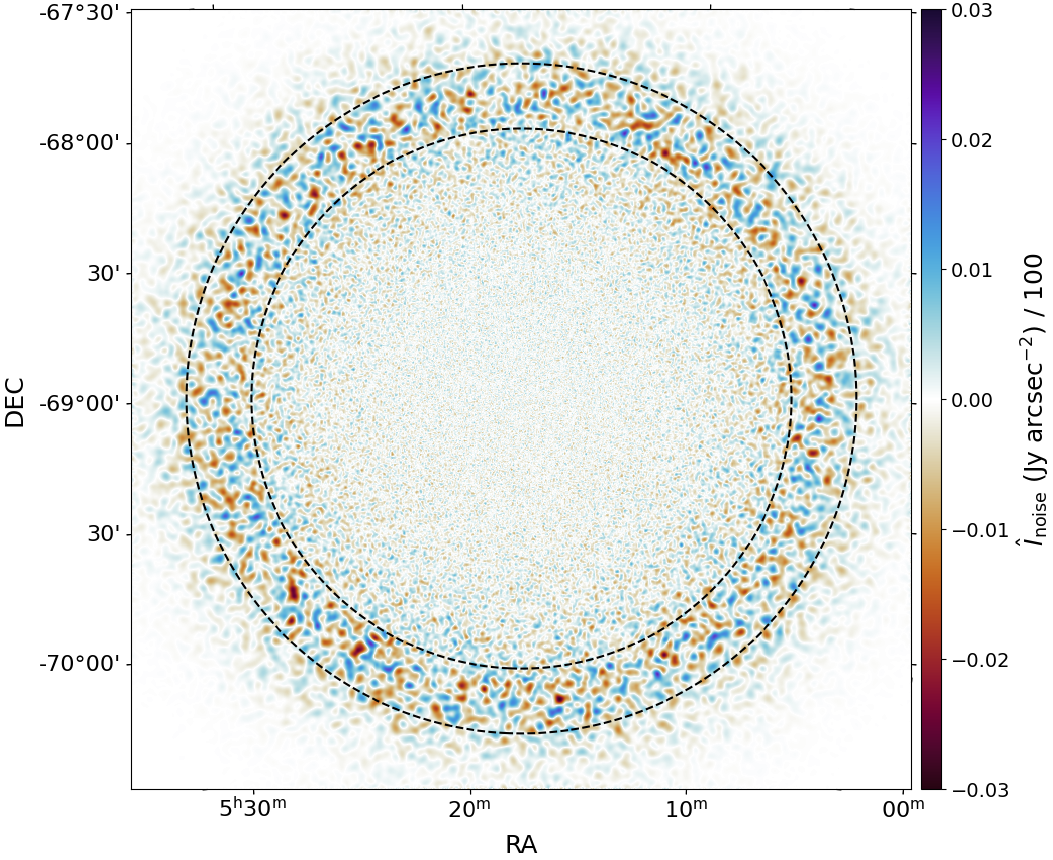}
  \caption{
    Recovered field $\hat\Ib'_k(\ell,m)$ from \CODE\, using \BASE\ with $\lambda_r=1$, of noise-only simulated visibilities.
    The black dashed contours show the primary beam values $\Ab_k(\ell,m)=0.01$ (outer) and $0.05$ (inner).
    }
  \label{fig:fbm_single_ivis_noise}
\end{figure}
In order to quantify the noise and effective beam response of the \BASE\ model, we simulated a single pointing observation, with a single spectral channel at 1420.4\,MHz, no astrophysical signal and a Gaussian random noise of 5\,Jy which was added independently to each complex visibility sample in {\tt CASA}.
The integration time was set to 500\,s, and we used the same spectral setup, polarization mode, and hour-angle coverage as in the point-source simulation.
However, the array configuration was chosen to match that of MeerKAT, with the pointing center set to $(\alpha,\delta) = (05^{\mathrm h}17^{\mathrm m}36^{\mathrm s}, -69^\circ02'00'')$, corresponding approximately to the center of the LMC.
By construction, the primary beam has no effect on the simulated visibilities because the product $\Ab_k\Ib_k'$ yields 0 if $\Ib_k'=0$. 
The reconstructed field on the other hand always includes the primary beam in the forward model used to fit the data and so the resulting image $\hat\Ib_k'$ will have response to the noise-only visibilities, but also to $\Ab_k$. 
Having $\Ab_k$ in the forward model also implies that there will be a natural coupling with the regularization term; Attenuation of the sky model $\Ib_k'$ by the primary beam will effectively modify the balance in the cost function between the data fidelity term $Q(\Ib)$ and the regularization term $R(\Ib)$. 
The response of \BASE\ to simulated noise is non trivial and we want to quantify it. 

Figure~\ref{fig:fbm_single_ivis_noise} shows the imaged field $\hat\Ib_k'$ from \CODE\, using \BASE\ with $\lambda_r=1$ and a pixel size of $14''$, with the black dashed contours showing the primary beam values $\Ab_k=0.01$ (outer) and $0.05$ (inner).
We find that the noise amplitude increases away from the pointing center, and that this enhanced noise is also characterized by larger spatial scales. This behavior is observed both with and without spatial regularization, suggesting that it is not solely driven by the regularization but also by the primary beam.
\begin{figure}
    \includegraphics[width=\linewidth]{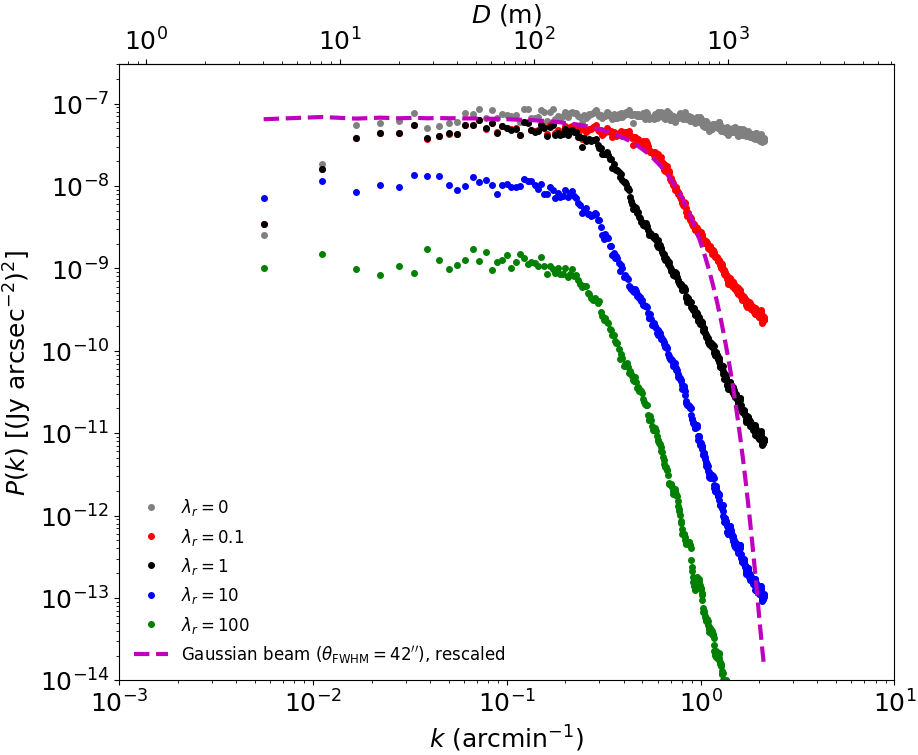}
  \caption{
    Power spectra $P(k)$ of the recovered field  $\hat\Ib_k'(\ell,m)$ from \CODE\, (\BASE) from a single-pointing MeerKAT simulation of interferometric noise only, for different value of $\lambda_r=(0,0.1,1,10,100)$.
    The top axis shows the equivalent projected baseline separation $D$.
    The magenta dashed line show a rescaled (at the mean level of the red curve) Gaussian beam with $\theta_{\mathrm{FWHM}}=42''$ (or three times the pixel size).
    }
  \label{fig:SPS_fbm_single_noise}
\end{figure}
We made use of the spatial power spectrum $P(k)$ to quantify how power is distributed as a function of scale across the entire image following the apodization procedure described in \citet[][see their section~3.1]{Marchal:2021}.
Figure~\ref{fig:SPS_fbm_single_noise} shows $P(k)$ of $\hat\Ib_k'$ for $\lambda_r=(0,0.1,1,10,100)$, shown in gray, red, black, blue, and green respectively.
In the absence of regularization, the power spectrum exhibits a relatively constant noise level at intermediate spatial scales. 
As $\lambda_r$ increases, the average noise level decreases and, as expected, high spatial frequencies are progressively suppressed.
This illustrates the coupling between noise suppression and the effective beam response: larger values of $\lambda_r$ reduce the apparent noise at the expense of a loss of effective resolution.

For reference, we also show a Gaussian beam with $\theta_{\mathrm{FWHM}} = 42''$, corresponding to approximately three times the pixel size of the grid (i.e., close to Nyquist sampling), and broadly consistent with the synthesized beam expected from the simulated MeerKAT observing setup at this declination.
We manually normalized the amplitude by selecting the value of $\lambda_r$ that best reproduces the inflection point toward higher spatial frequencies, and find that $\lambda_r = 0.1$ yields an effective beam response closely approximated by this Gaussian beam.
Other Gaussian beams with different $\theta_{\mathrm{FWHM}}$ could be matched to each value of $\lambda_r$; in practice, this provides a convenient way to quantify the effective beam response of \CODE\ for a given regularization strength.
Note however that this does not quantify the spatial variation of the effective resolution across the field. In practice, the achievable effective resolution is expected to gradually degrade with increasing distance from the pointing center, as the primary beam attenuates the signal and therefore reduces the sensitivity to high spatial frequencies, even in the absence of explicit regularization ($\lambda_r = 0$).
To investigate this effect in more detail, we analyzed the radial variation of the effective resolution across the reconstructed field.
 
\subsection{Radial Variation of the Effective Resolution}
\label{sec:radial_resolution}
\begin{figure}
    \includegraphics[width=\linewidth]{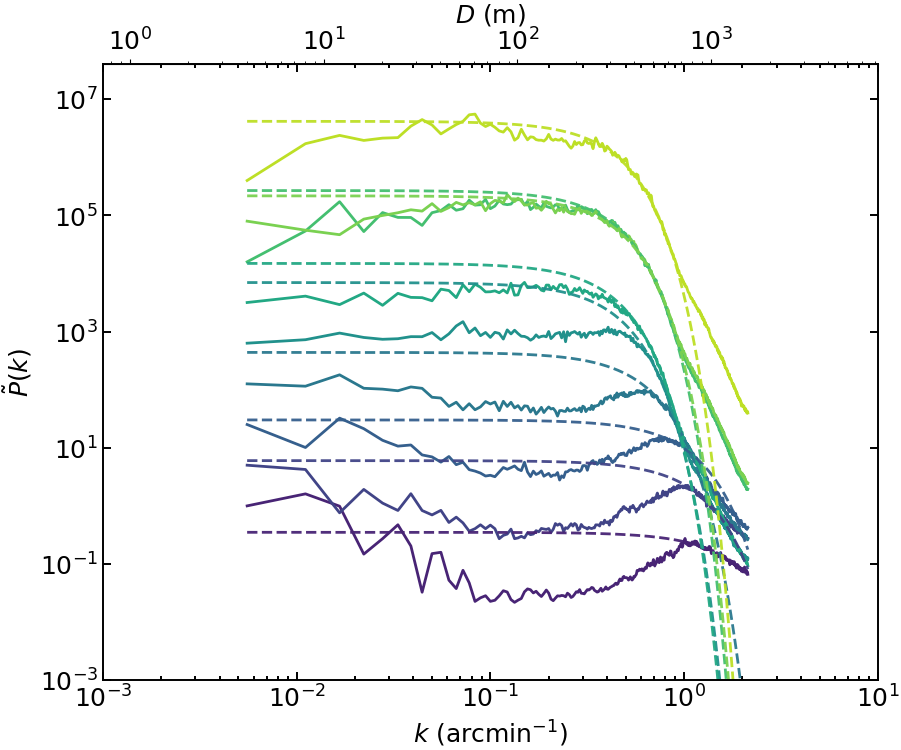}
  \caption{
    Power spectra $\tilde P(k)$ measured in concentric annuli centered on the pointing center.
    The top axis shows the equivalent projected baseline separation $D$.
    The amplitudes have been normalized and arbitrarily shifted for visualization purposes. 
    Each curve corresponds to a different projected angular distance $\delta$ from the field center. 
    The dashed lines show the corresponding Gaussian beam fits used to estimate the effective angular resolution in each annulus.
    }
  \label{fig:varying_res_annuli_sps}
\end{figure}
\begin{figure}
    \includegraphics[width=\linewidth]{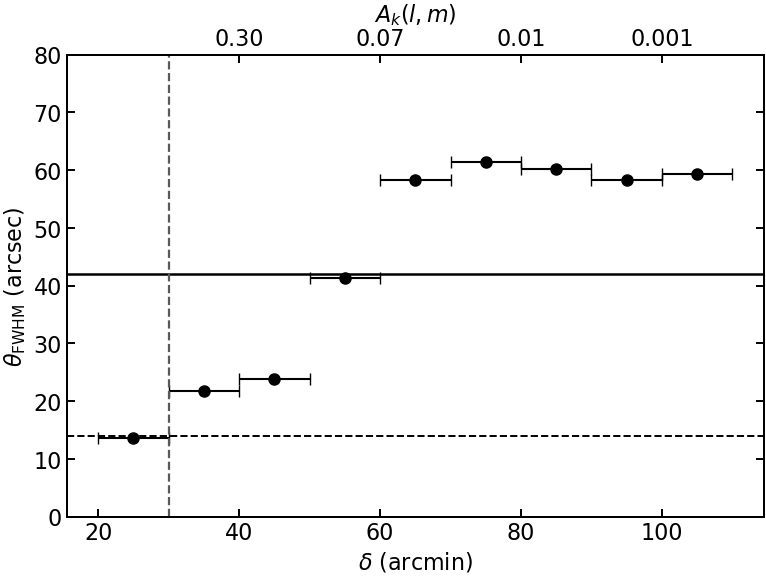}
  \caption{
    Effective angular resolution as a function of projected angular distance $\delta$ from the pointing center.
    The top axis shows to corresponding primary beam level values. 
    The values of $\theta_{\rm FWHM}$ were estimated from the beam fits shown in Figure~\ref{fig:varying_res_annuli_sps}. 
    Horizontal error bars indicate the radial extent of each annulus used.
    The horizontal solid line indicates $\theta_{\mathrm{FWHM}}=42''$ (or three times the pixel size), corresponding to the magenta dashed line in Figure~\ref{fig:SPS_fbm_single_noise}.
    The horizontal dashed line indicates the pixel size of the grid.
    The vertical dotted line indicates the half-power radius of the primary beam, $\mathrm{HPBW}/2$}.
  \label{fig:varying_res_annuli_sps_resolution}
\end{figure}
\begin{figure}
    \includegraphics[width=\linewidth]{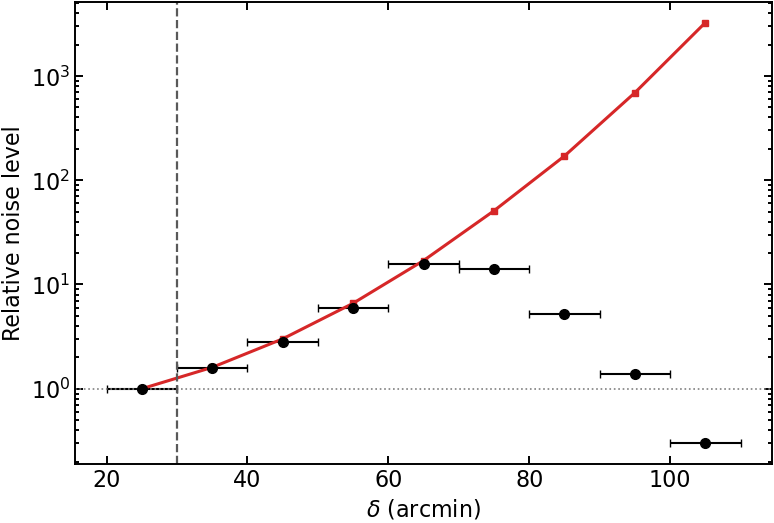}
  \caption{
    Relative small-scale noise level as a function of projected angular distance \(\delta\) from the pointing center. 
    The reconstructed noise level is estimated from the square root of the approximately flat power-spectrum amplitude over \(0.1 \leq k \leq 0.5~{\rm arcmin}^{-1}\), corrected for the effective area of each annular window and normalized to the innermost annulus. 
    The red curve shows the $\Ab(\delta)^{-1}$ scaling expected for the noise in a primary-beam-corrected CLEAN image. 
    Horizontal error bars indicate the radial extent of each annulus.
    The vertical dotted line indicates the half-power radius of the primary beam, $\mathrm{HPBW}/2$.
    }
  \label{fig:varying_res_annuli_noise}
\end{figure}
We measured the spatial power spectrum of the reconstructed image in a series of concentric annuli defined in real space (i.e., in the image plane) and centered on the pointing center. 
Each annulus was defined using a radially tapered mask $w_{\rm ann}(\rb)$, which smoothly transitions between zero and unity at its inner and outer boundaries.
For each annulus, the azimuthally averaged power spectrum $\tilde P(k)$ was computed independently. By comparing the resulting spectra at different projected angular distances from the pointing center, we can probe how the effective beam response varies across the field.
This analysis was performed on the same noise-only simulation discussed in Section~\ref{subsec:noise}, using the $\lambda_r=0.1$ reconstruction corresponding to the red curve in Figure~\ref{fig:SPS_fbm_single_noise}.
The characteristic beam scale in each region was estimated by fitting the spectrum in the vicinity of its inflection point, where the beam-induced turnover becomes apparent.
Figure~\ref{fig:varying_res_annuli_sps} shows the resulting spatial power spectra measured in the different annuli. The dashed lines show the associated Gaussian beam fits used to estimate the effective angular resolution in each annulus. The progressive shift of the turnover toward lower spatial frequencies with increasing $\delta$ indicates a gradual degradation of the effective angular resolution away from the pointing center.
The derived values of $\theta_{\rm FWHM}$ are shown in Figure~\ref{fig:varying_res_annuli_sps_resolution} as a function of $\delta$, with horizontal error bars indicating the radial extent of each annulus. 
The vertical dotted line indicates the half-power radius of the primary beam, corresponding to half the half-power beam width ($\mathrm{HPBW}/2$).
This provides an empirical characterization of the radial evolution of the effective resolution across the reconstructed field.

We also used these annuli to quantify the radial variation of the reconstructed noise and compare it with the behavior expected for a primary-beam-corrected {\tt CLEAN} image.
For each annulus, we estimated the small-scale noise level from the approximately flat power-spectrum amplitude $P_{\rm flat}$ over $0.1 \leq k \leq 0.5~{\rm arcmin}^{-1}$ as $\sigma_{\rm noise} \propto [P_{\rm flat}/\langle w_{\rm ann}^2\rangle]^{1/2}$, where the $\langle w_{\rm ann}^2\rangle$ normalization accounts for the different effective areas of the tapered annuli.
Figure~\ref{fig:varying_res_annuli_noise} compares the resulting noise levels, normalized to the innermost annulus, with the $\Ab(\delta)^{-1}$ scaling expected from primary-beam correction.
The reconstructed noise closely follows the $\Ab(\delta)^{-1}$ scaling over the inner part of the primary beam.
At larger distances from the pointing center, however, the small-scale noise departs strongly from this relation and eventually decreases.
This does not correspond to a reduction of the overall reconstruction noise, but rather reflects the increasingly correlated, scale-dependent noise seen in Figure~\ref{fig:varying_res_annuli_sps}, for which power is progressively shifted toward larger spatial scales.

In practice, however, for the joint-deconvolution mosaics presented below, the combination of multiple overlapping pointings yields a much more uniform sensitivity pattern across the field, and therefore a more spatially uniform effective resolution and noise level.
These results illustrate that, within the \CODE\ framework, the effective beam response is not solely determined by the interferometric sampling function, but also emerges from the interplay between primary beam attenuation, noise statistics, and regularization.
As in traditional {\tt CLEAN}-based approaches, one may convolve the reconstructed solution with a restoring beam in order to obtain a well-characterized effective resolution. The key difference here is that the regularization modifies the reconstructed model itself, such that the noise suppressed during the optimization is naturally incorporated into the effective beam response rather than being reintroduced through a residual map.
Finally, this characterization could be extended to a two-dimensional power spectrum, allowing anisotropies in the effective beam response to be quantified directly in Fourier space, for example using an elliptical Gaussian model with a position angle and two $\theta_{\mathrm{FWHM}}$ values.
Such an extension would be particularly relevant for observations with highly anisotropic $uv$ coverage (e.g., VLA observations).

\subsection{Synthetic observations of a fBm} \label{subsec:synthetic}
\begin{figure}
    \includegraphics[width=\linewidth]{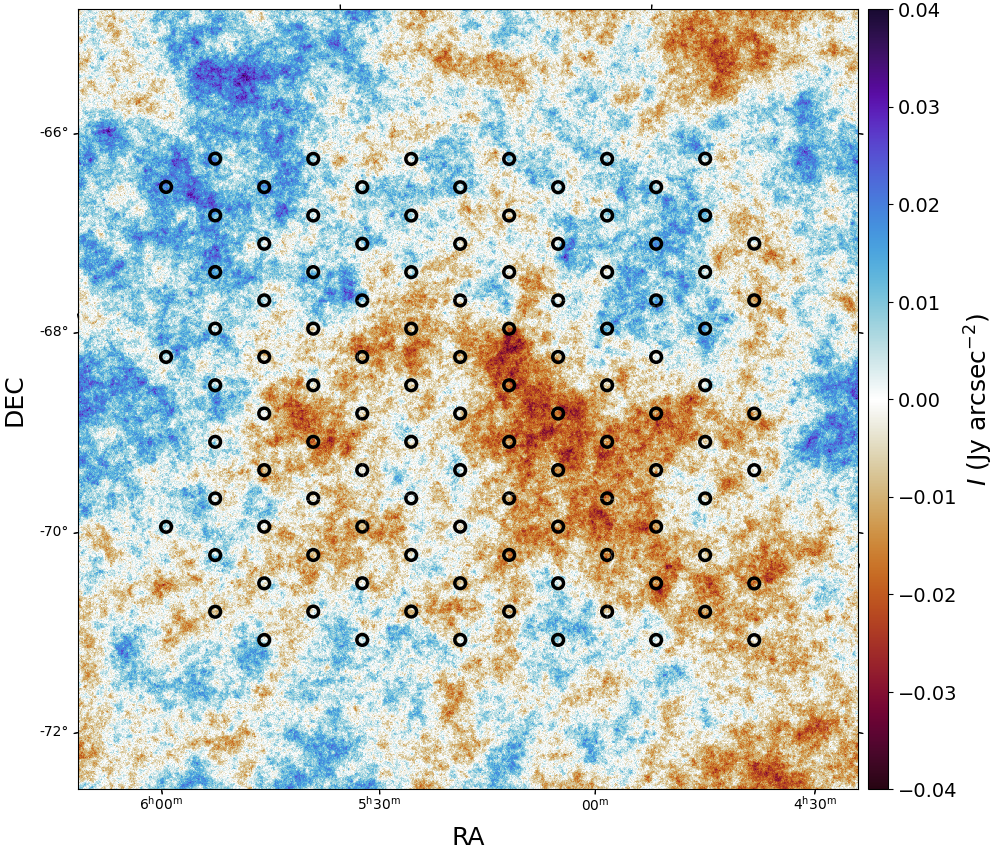}
  \caption{
    Fractional Brownian motion field with spatial power spectrum slope index $n=2.4$, with size 2048$^2$ and pixel size of $14''$, centered at $(\alpha,\delta) = (05^{\mathrm h}17^{\mathrm m}36^{\mathrm s},-69^\circ02'00'')$.
    The black circle show the location of the ASKAP-like mosaic with 108 pointings. 
    }
  \label{fig:fbm_single_ivis_pointings}
\end{figure}

We generated an idealized simulation of multiscale diffuse emission using the \texttt{fBmnd} package\footnote{\url{https://github.com/antoinemarchal/fBms}}. The model consists of a two-dimensional fractional Brownian motion (hereafter fBm) image with $2048^2$ pixels of size $14''$, again centered at $(\alpha,\delta) = (05^{\mathrm h}17^{\mathrm m}36^{\mathrm s}, -69^\circ02'00'')$, corresponding approximately to the center of the LMC.
The emission is characterized by a single power-law spatial power spectrum with spectral index $n=2.4$. The spatial-frequency range was left unconstrained, producing a scale-free structure across all spatial modes in the image.

An ASKAP-like mosaic was constructed by tiling the fBm sky model using a three-interleave close-packed footprint consisting of 108 individual pointings arranged in a $6\times6$ hexagonal pattern per interleave \citep{Hotan:2021}, mimicking the Phased Array Feed (PAF) mounted on each of the 36 ASKAP antennas.
The beam centers were defined in the local SIN-projected tangent plane around the mosaic center, with a constant angular separation of $1^\circ$ between adjacent pointings, while successive rows were offset by half this separation to form a close-packed hexagonal geometry.
Three interleaves were generated by applying symmetric positional offsets forming an equilateral-triangle pattern, ensuring uniform spatial sampling.
Figure~\ref{fig:fbm_single_ivis_pointings} shows the simulated fBm field with the black circles annotating the positions of the 108 pointings.
For each pointing, a Gaussian primary beam $\Ab_k\equiv\Gb$ with $\theta_{\mathrm{FWHM}}=1^\circ$ was generated on an image grid extending to $1.5\times\theta_{\mathrm{FWHM}}$ from the beam center. The fBm sky image was then individually re-projected onto the corresponding SIN-projected primary beam grid.
Each pointing yielded an attenuated sky image $\Ab_k\Ib_k'$, written on disk in a FITS format and provided to CASA to simulate the 108 measurement sets.
Here again, we simulated a single spectral channel at 1420.4\,MHz.
Note that the simulated mosaic adopts an ASKAP-like pointing geometry, while the interferometric visibilities themselves were generated using a MeerKAT array configuration.
The integration time was set to 500\,s, and we used the same array configuration (i.e., with a MeerKAT layout), spectral setup, polarization mode, and hour-angle coverage as in the point-source simulation.
It yielded a total of about 24\,M visibilities, about 6 times the total number of pixels in the original field (i.e., 2048$^2$).
We also simulated (not shown here), a Parkes-like single dish observation obtained by convolving the fBm field with a Gaussian kernel with $\theta_{\mathrm{FWHM}}=16^\prime$.

\subsubsection{Linear mosaicking}
\begin{figure}
    \includegraphics[width=\linewidth]{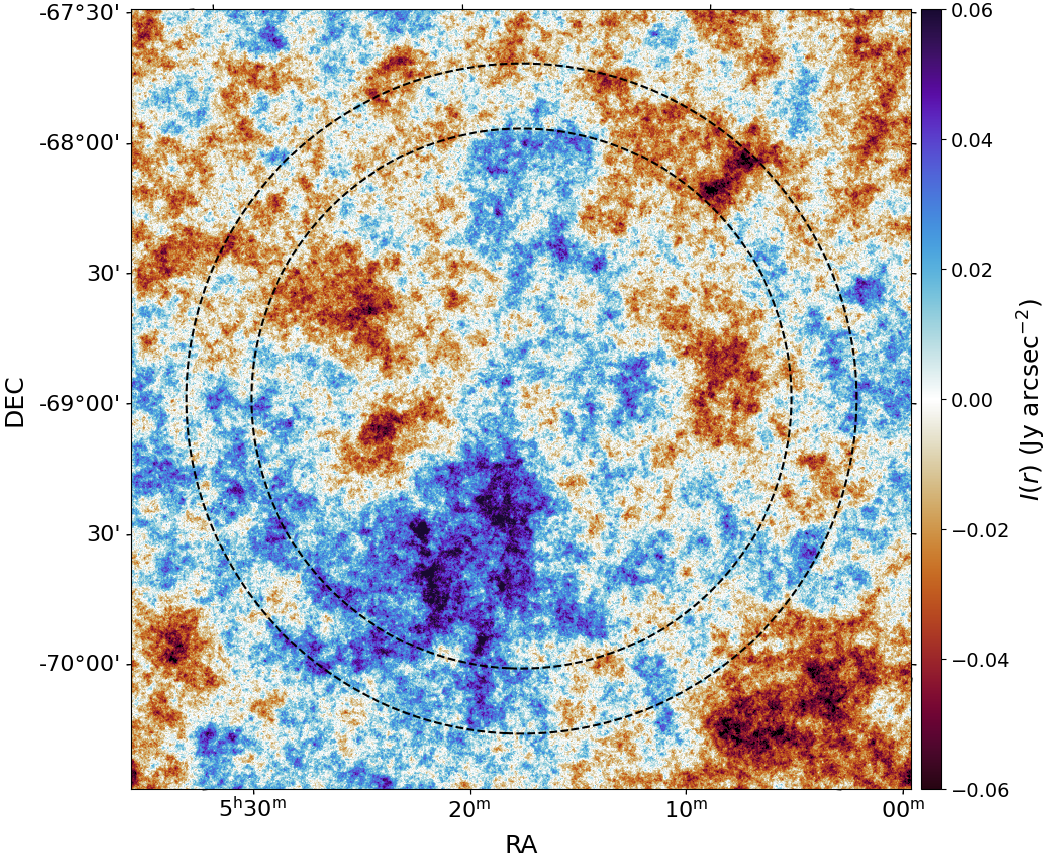}
    \includegraphics[width=\linewidth]{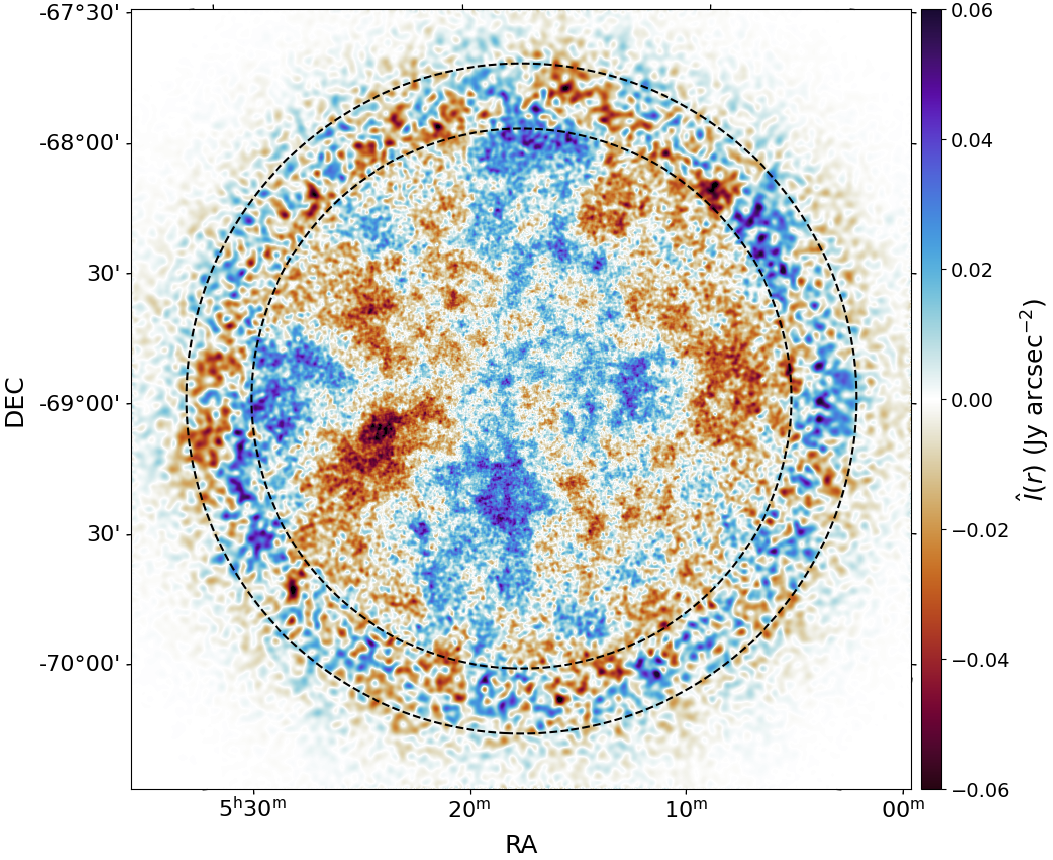}
  \caption{
    Top: Projected map $\Ib'_k(\ell,m)$ of a single pointing in the simulated ASKAP-like mosaic from Figure~\ref{fig:fbm_single_ivis_pointings}. 
    Annotations are as in Figure~\ref{fig:fbm_single_ivis_noise}.
    Bottom: Recovered field $\hat\Ib'_k(\ell,m)$ from \CODE\, using \BASE. 
    }
  \label{fig:fbm_single_true}
\end{figure}
\begin{figure*}
    \includegraphics[width=0.49\linewidth]{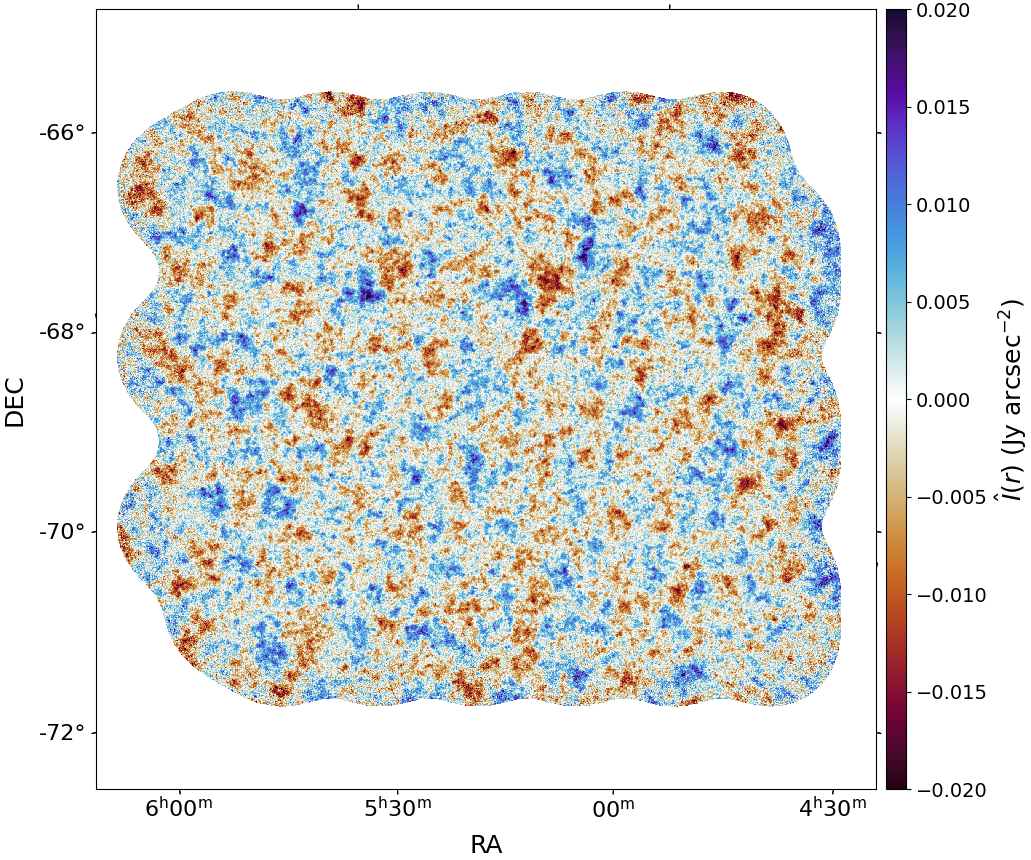} 
    \includegraphics[width=0.49\linewidth]{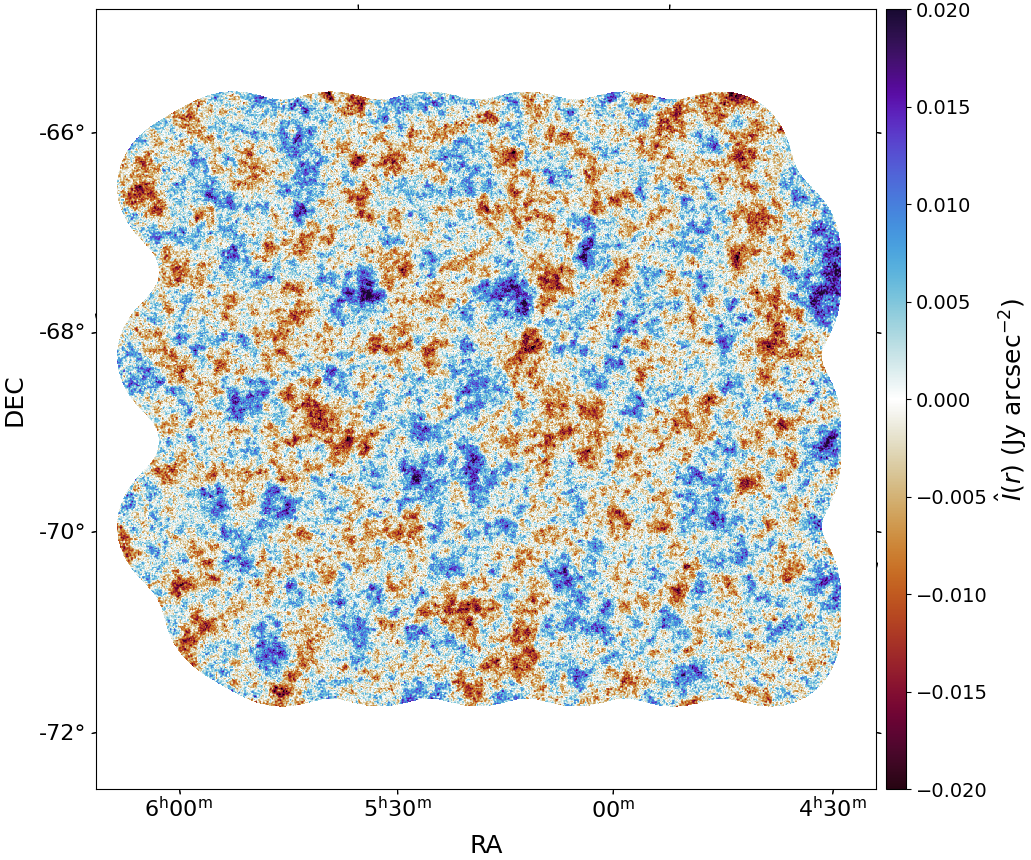} \\
    \includegraphics[width=0.49\linewidth]{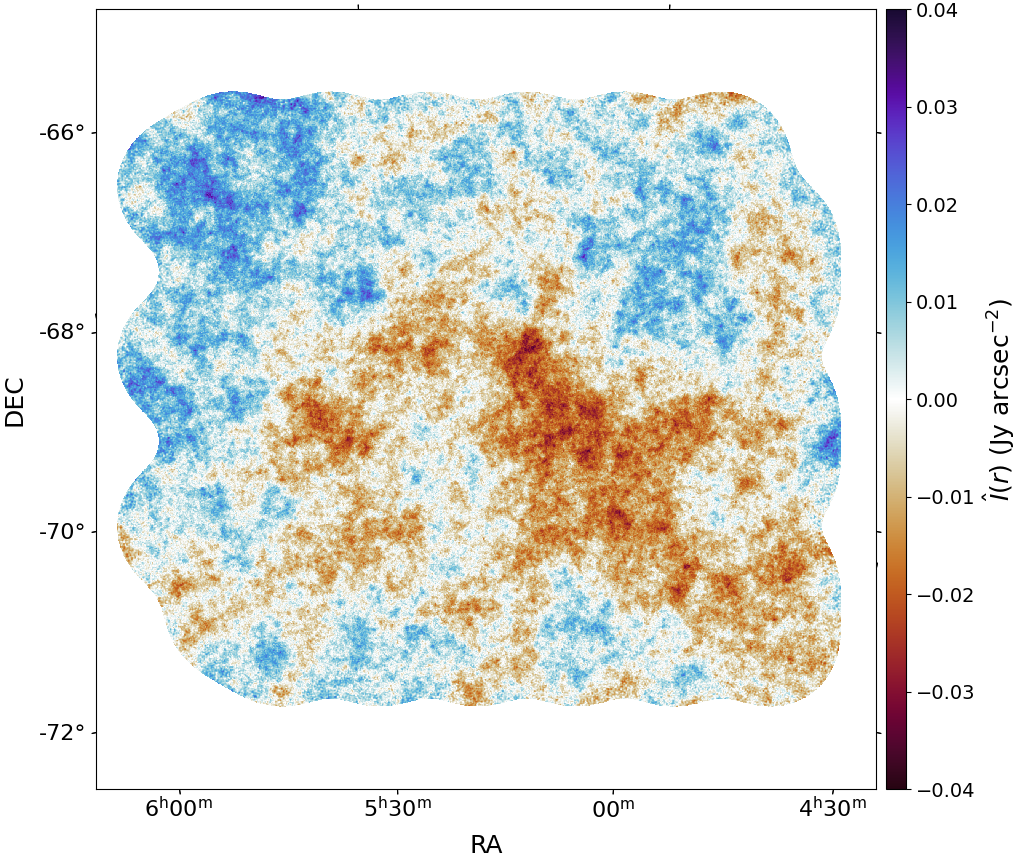}
    \includegraphics[width=0.49\linewidth]{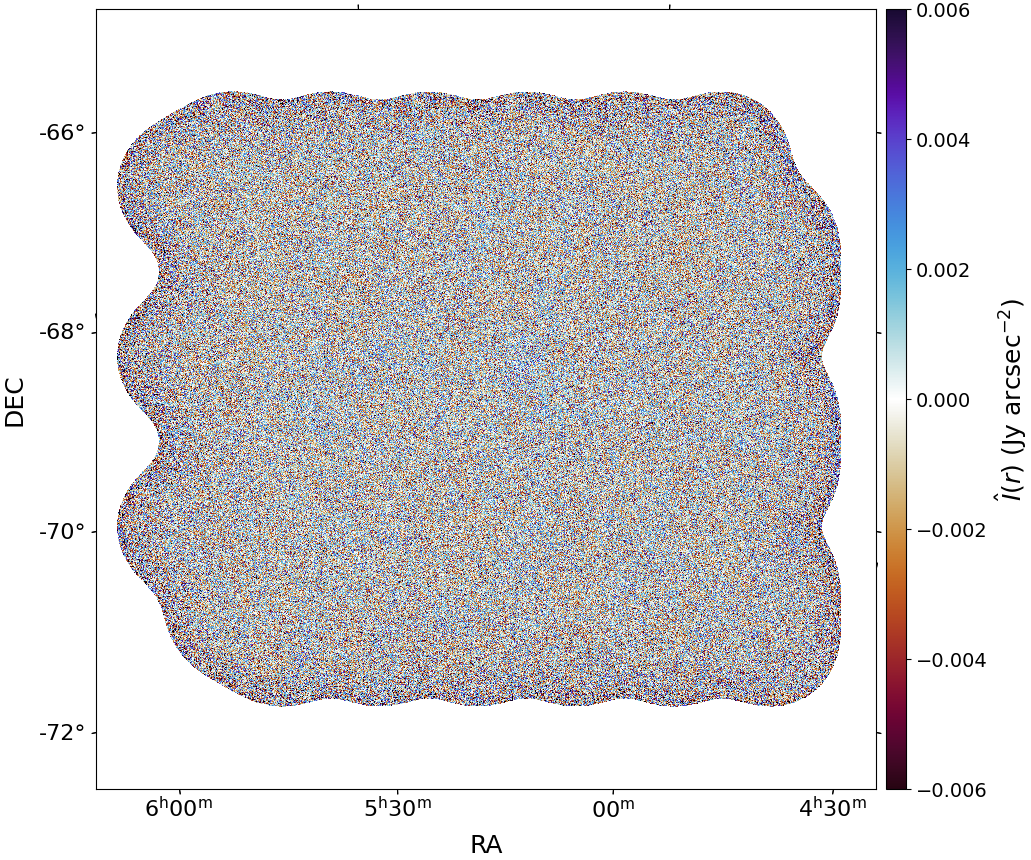}
  \caption{
    Recovered sky model $\hat\Ib_{\nu}(\rb)$ of the simulated ASKAP-like mosaic for linear mosaicking (top left), joint deconvolution (top right), joint deconvolution with fusion of single-dish data (bottom left), and its residual with the simulated fBm field (Figure~\ref{fig:fbm_single_ivis_pointings}, top).
    We applied a mask corresponding to an effective primary-beam level of 0.2.
    }
  \label{fig:fbm_linear_mosaic_ivis}
\end{figure*}
We selected the central pointing from the mosaic and imaged it with \CODE\ (\BASE), with $\lambda_r=1$ and no positivity constraint.
A local projection toward the pointing center is shown in the top panel of Figure~\ref{fig:fbm_single_true}, with the recovered sky model $\hat\Ib'_k$ displayed in the bottom panel.
We find that the solution is visually consistent with the input field near the image center, while the correlation becomes progressively weaker toward larger angular distances from the pointing center.
This behavior is expected from the attenuation of the primary beam.
Additionally, $\hat \Ib'_k$ shows reduced power on large spatial scales compared to $\Ib'_k$, which is also consistent with the absence (or poor sampling) of short baselines.
We repeated the deconvolution for each pointing independently and combined them using a linear mosaicking (used here only as a reference case without joint deconvolution; this should not be confused with the joint-deconvolution method of \citet{Sault:1996}, which operates on a linearly mosaicked dirty image).
We first re-projected each sub-image
\begin{align} \label{eq:sin_deprojection}
    \hat \Ib'_{k,\nu}(\rb)
    \xleftarrow{\textbf{projection}}
    \hat \Ib'_{k,\nu}(\ell,m) \, ,
\end{align}
and linearly combined them such as
\begin{align} \label{eq:linear}
    \hat \Ib_{\nu}(\rb) =
    \frac{\sum_k \Ab_{k,\nu}^2(\rb)\,\hat \Ib'_{k,\nu}(\rb)}
    {\sum_k \Ab_{k,\nu}^2(\rb)} \, .
\end{align}
The corresponding effective primary beam is
\begin{align} \label{eq:pb_eff}
    \Ab^{\rm eff}_{\nu}(\rb) =
    \sqrt{\sum_k \Ab_{k,\nu}^2(\rb)} \, .
\end{align}
\begin{figure}
    \includegraphics[width=\linewidth]{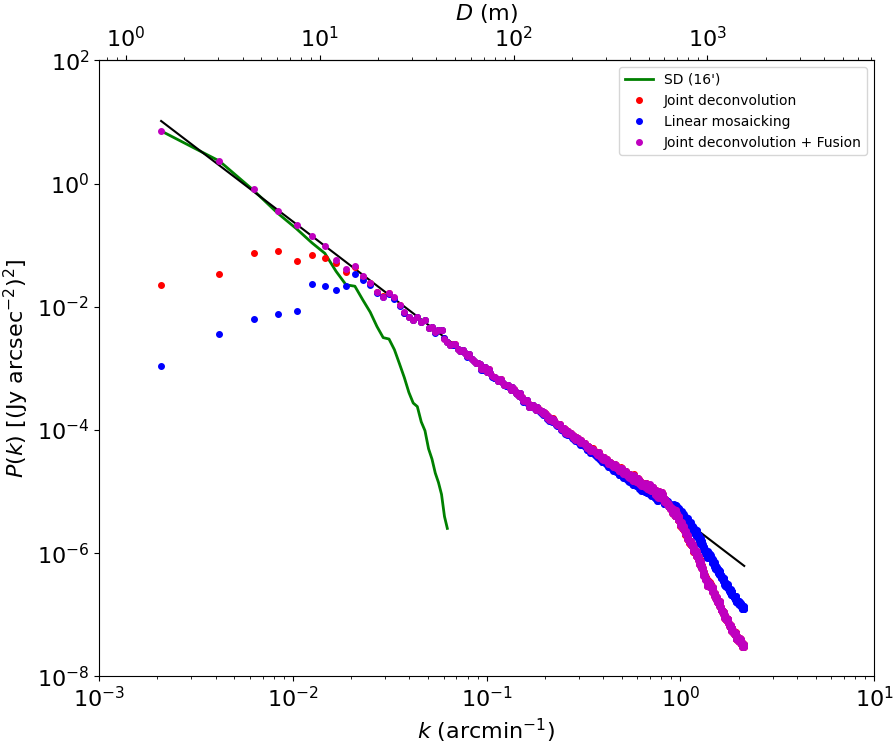}
    \includegraphics[width=\linewidth]{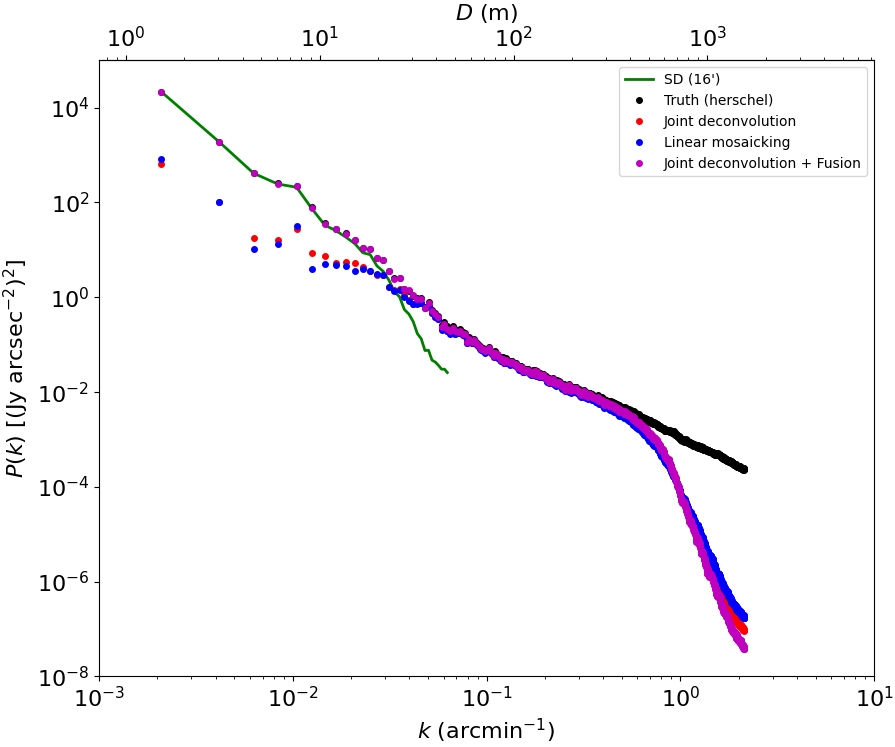}
  \caption{
    Power spectra $P(k)$ from pointings of the simulated ASKAP-like mosaic.
    The top axis shows the equivalent projected baseline separation $D$.
    The top panel shows the Fractional Brownian motion simulated field.
    The bottom panel shows the mock intensity map from the \Herschel\, $250\,\mu\mathrm{m}$ observation of the Spider region.
    The black line/points shows the input simulated field.
    Blue points show the recovered field using linear mosaicking of individually deconvolved pointings. 
    Red points are from the joint deconvolution performed with \CODE\, (\BASE) with $\lambda_r=1$. 
    Magenta points are from the joint deconvolution with fusion of single-dish data.
    The green line shows the simulated Parkes-like observation.
    }
  \label{fig:SPS_fbm_mosaic}
\end{figure}
The result is shown in the top left panel of Figure~\ref{fig:fbm_linear_mosaic_ivis}, where we apply a mask corresponding to an effective primary-beam level of 0.2.
The suppression of low spatial frequencies, relative to the simulated fBm (Figure~\ref{fig:fbm_single_ivis_pointings}), is clearly visible.
This can be appreciated in Figure~\ref{fig:SPS_fbm_mosaic} (top panel) which shows the corresponding spatial power spectrum in blue, compared to the solid black line representing the input fBm power law.
$P(k)$ of the recovered field departs sharply from that of the simulation at $k = 1.5\times10^{-2}$\,arcmin$^{-1}$, or an equivalent projected baseline separation $D$ of about $20$\,m (top label). 

\subsubsection{Joint deconvolution}
\label{subsec:joint-deconv}
The joint deconvolution of all the pointings was performed using the exact same parameters: number of iterations and hyper-parameters, and we show the result in the top right panel of Figure~\ref{fig:fbm_linear_mosaic_ivis}. 
We find that, in this case, the joint deconvolution exhibits a larger power at low spatial frequencies, which compared to the linear mosaicking can be appreciated both visually and in the spatial power spectrum (red; top panel in Figure~\ref{fig:SPS_fbm_mosaic}).
Although deviations from the input power law remain, the reconstructed spectrum follows it more closely, up to a projected baseline of about $10$\,m.
This behavior is consistent with the findings of \citet{Cornwell:1988}, the seminal study establishing joint deconvolution as a framework for synthesis imaging of large diffuse objects.
Finally, we imaged the data using the joint deconvolution with fusion ($\lambda_s=\lambda_r=1$), making use of the Parkes-like synthetic observation at 16$^\prime$, and the result is shown in the bottom left panel in Figure~\ref{fig:fbm_linear_mosaic_ivis}.
The residual between this model and the simulated fBm is shown in the bottom right panel.
The power spectrum of the single-dish data is shown by the solid green line in Figure~\ref{fig:SPS_fbm_mosaic}, while that of the imaged field is shown in magenta.
We find a good agreement, both visually and quantitatively.

\subsection{Synthetic observation with a mock intensity map from \Herschel}
\label{subsec:herschel}
\begin{figure}
    \includegraphics[width=\linewidth]{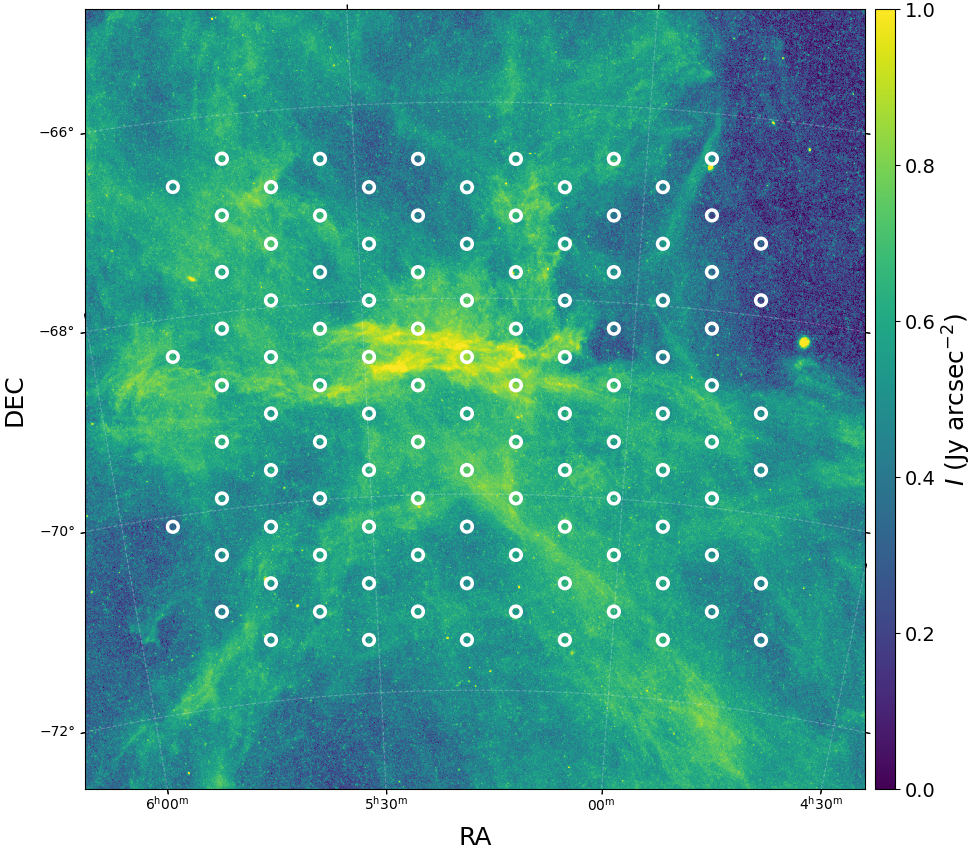}
    \caption{
    Mock intensity map constructed from a \Herschel\, $250\,\mu\mathrm{m}$ observation of the Spider region and re-projected toward $(\alpha,\delta) = (05^{\mathrm h}17^{\mathrm m}36^{\mathrm s},-69^\circ02'00'')$ to match the simulated ASKAP-like mosaic used in this work.
    The map is used as a realistic diffuse-emission sky model to generate synthetic interferometric observations.
    The mock radio intensity is scaled between 0 and 1\,Jy\,arcsec$^{-1}$.
    Annotations are as in Figure~\ref{fig:fbm_single_ivis_pointings}.
    }
  \label{fig:herschel_single_ivis_pointings}
\end{figure}
Using an fBm simulation provides a first benchmark to quantify the fidelity of the recovered deconvolved emission, but it does not capture the richness and complexity of the multiphase, magnetized interstellar medium (ISM) revealed by the various radio emission lines.
This complexity is statistically characterized by non-Gaussianity and anisotropy, and couplings between spatial scales \citep{Bruna:2013,Allys:2019}; features that are intrinsically absent from fBm realizations used here.
To introduce greater realism, we turned to far-infrared observations of the ISM from \Herschel, specifically the Spider region located at the top of the North Celestial Pole Loop \citep{Marchal:2023,Zhang:2023}.
Although this region is not observable with MeerKAT, we retained for practical purposes the same pointing directions as in the previous experiment (toward the LMC), as well as the exact same observing setup (including pointing directions that are typical of ASKAP's PAF).
The pixel size of the original \Herschel\, data does not match that of the grid used here ($14''$); consequently, the Spider region is not shown to scale.
The goal is simply to provide a more realistic mock sky model, shown in Figure~\ref{fig:herschel_single_ivis_pointings}, from which to generate synthetic interferometric observations.
While the \Herschel\, image contains bright compact structures, these are already convolved with the instrumental PSF and therefore do not constitute truly unresolved point sources that an interferometer would see. 
This experiment should therefore not be interpreted as a point-source plus diffuse-emission validation test, but rather as a test of the recovery of structured diffuse emission spanning a broad range of angular scales.
Note also that the mock radio intensity was scaled between 0 and 1\,Jy\,arcsec$^{-1}$ to yield a positive-only sky model, before the visibilities were generated with \texttt{CASA}.

\begin{figure*}
    \includegraphics[width=0.49\linewidth]{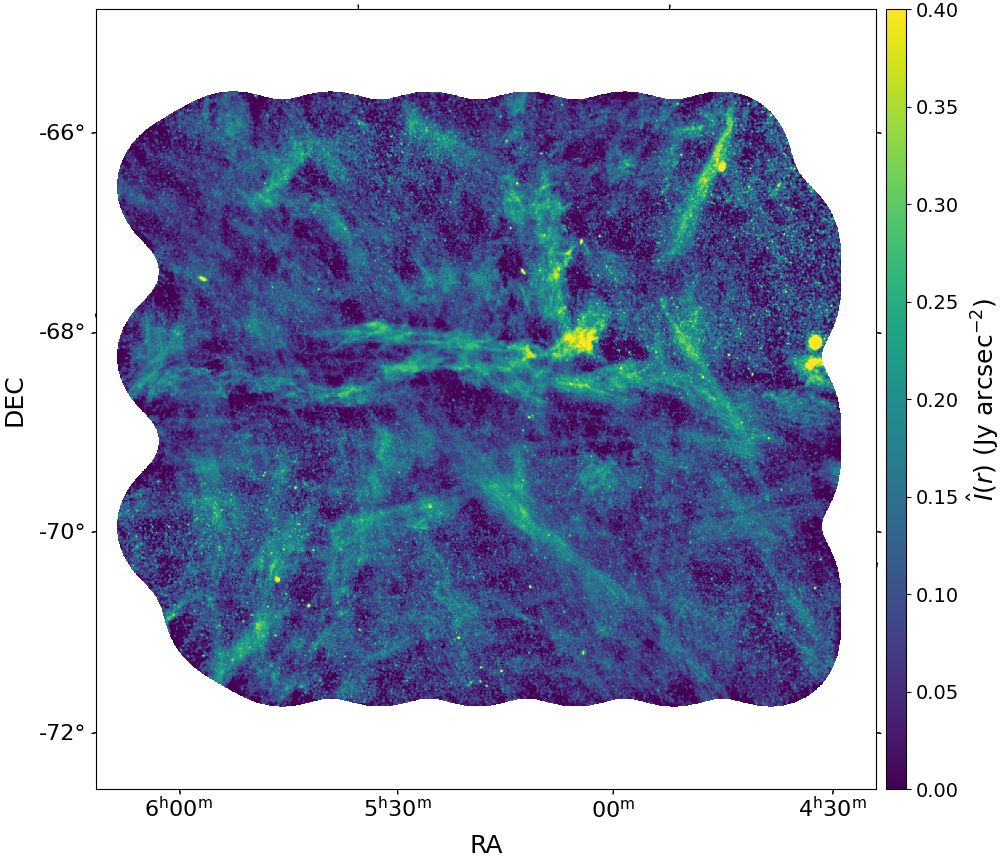}
    \includegraphics[width=0.49\linewidth]{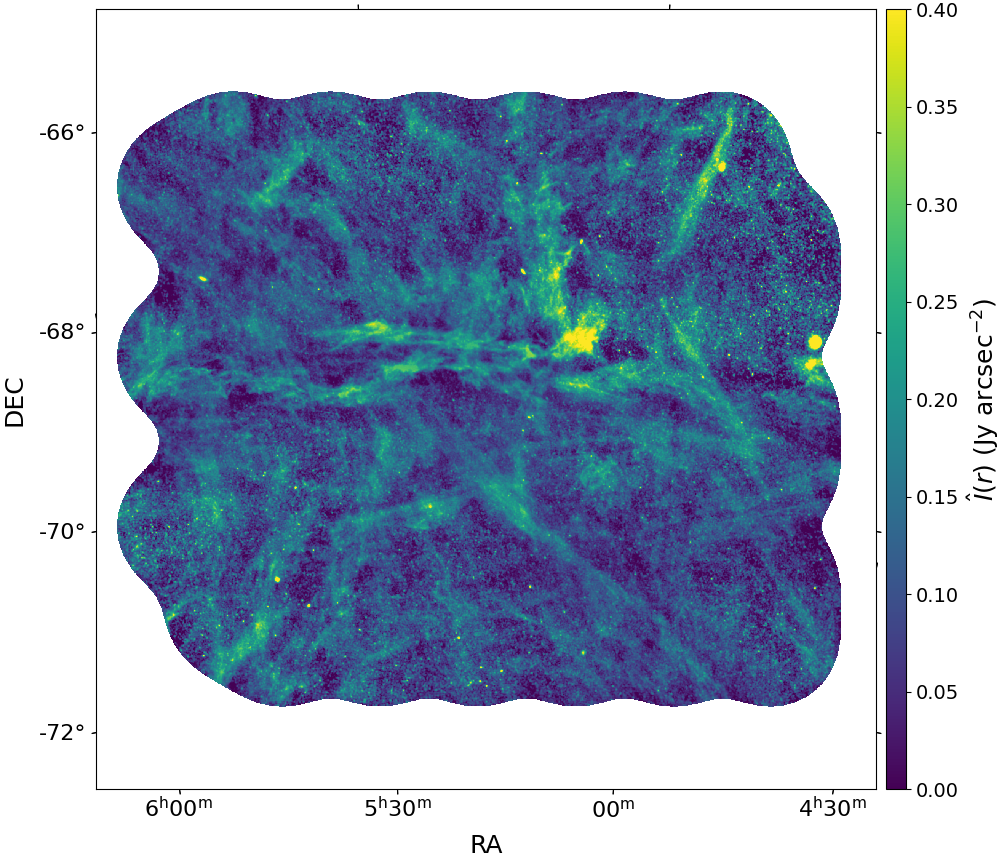}
    \includegraphics[width=0.49\linewidth]{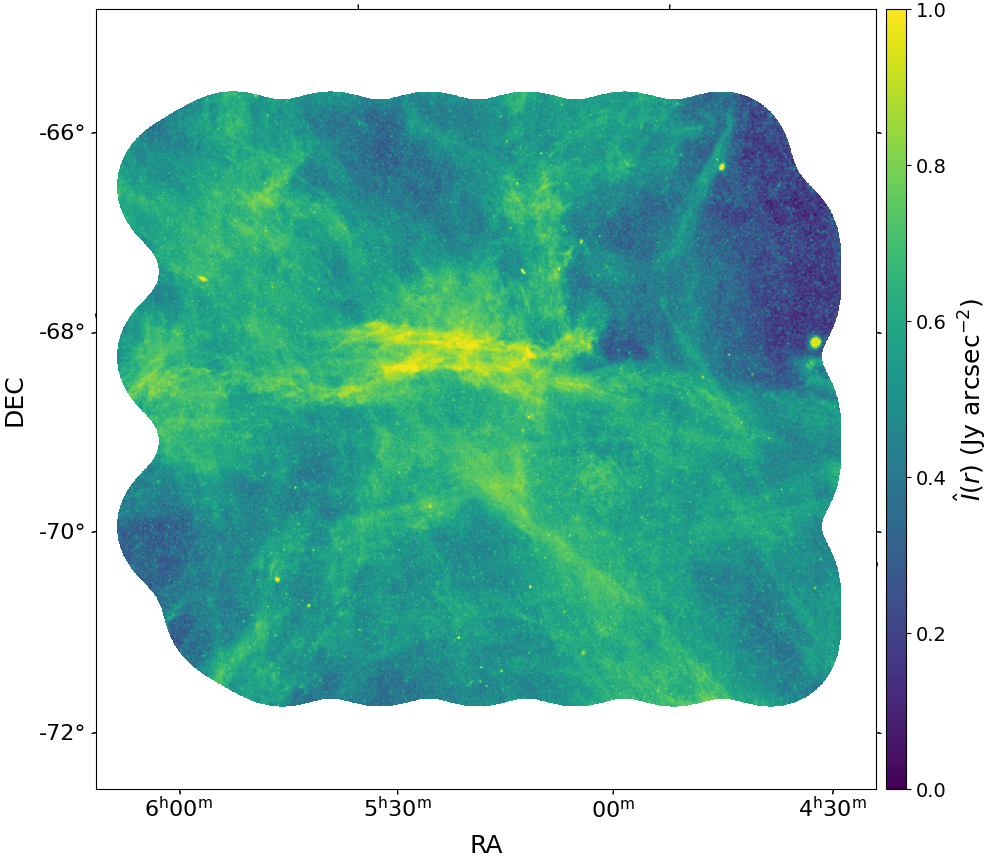}    \includegraphics[width=0.49\linewidth]{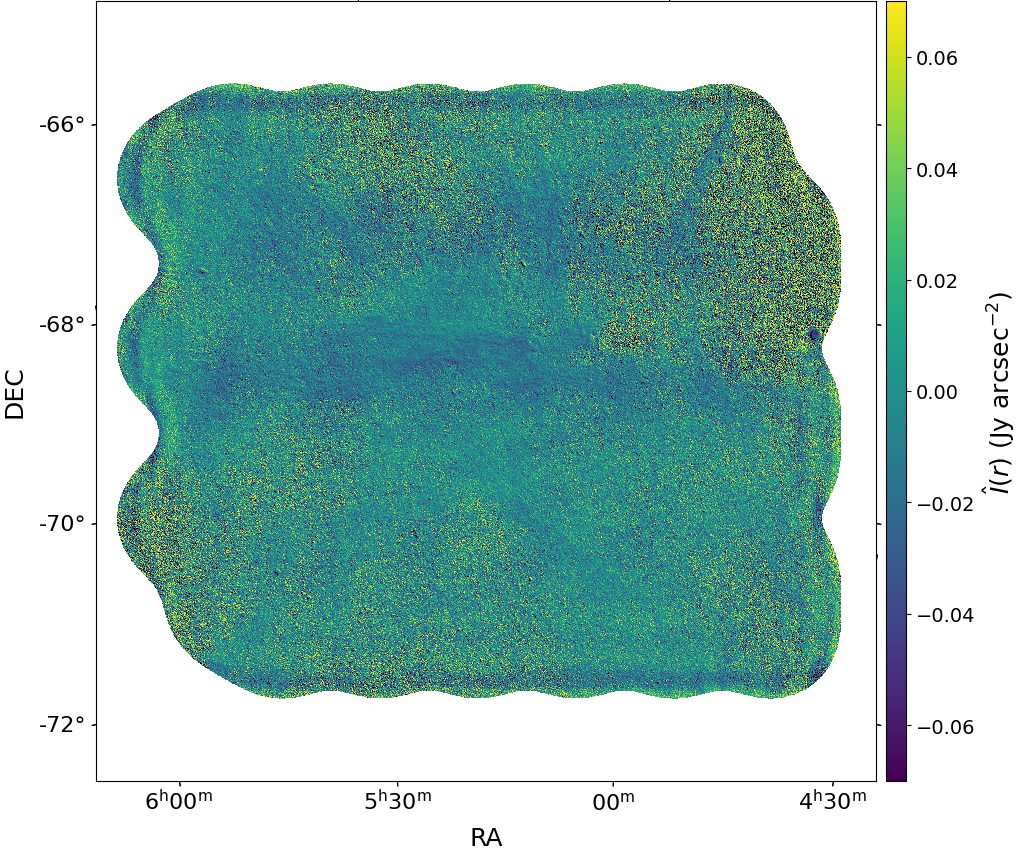}
  \caption{
    Same as Figure~\ref{fig:fbm_linear_mosaic_ivis} but for the mock intensity map from the \Herschel\, $250\,\mu\mathrm{m}$ observation of the Spider region.
    }
  \label{fig:herschel_mosaic_ivis}
\end{figure*}
\begin{figure}
    \includegraphics[width=\linewidth]{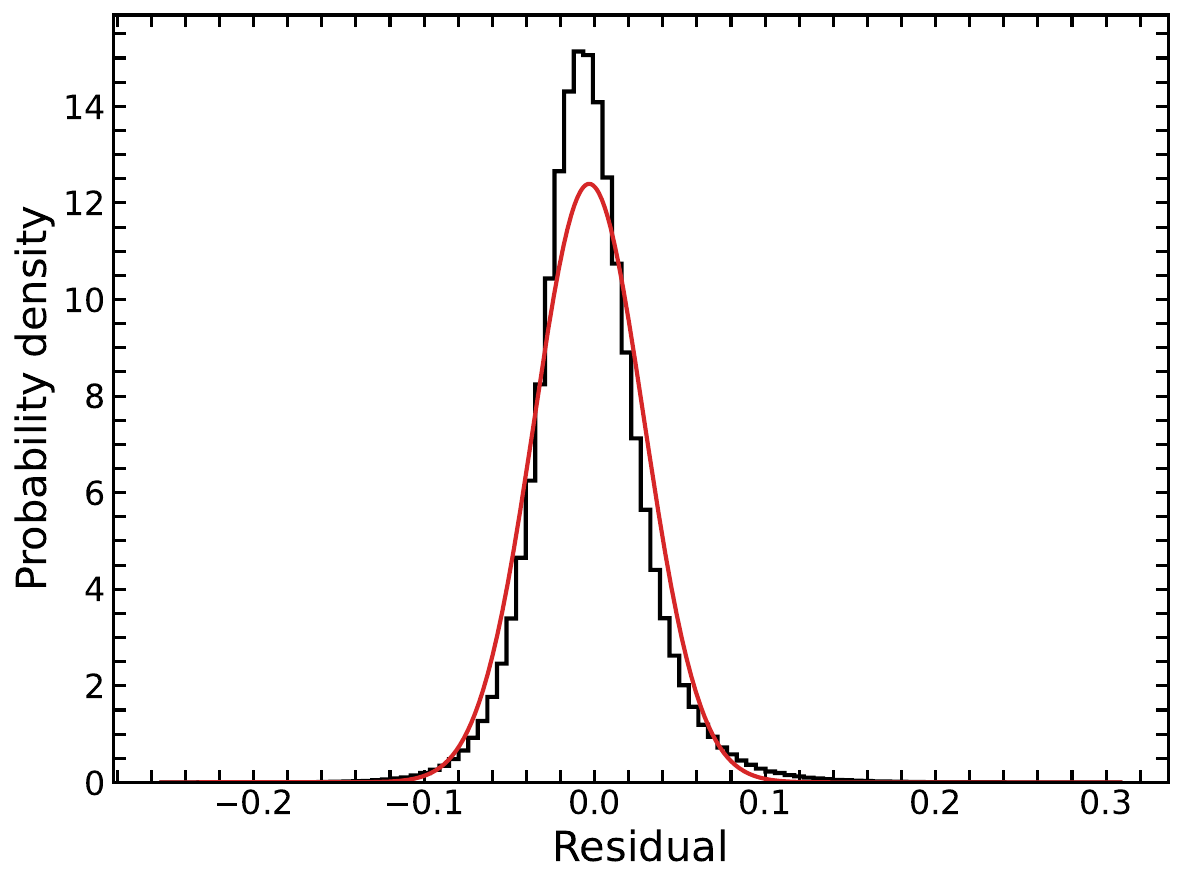}
    \caption{
    Normalized histogram of the residual map, for the Spider \texttt{Herschel} field, shown in Figure~\ref{fig:herschel_mosaic_ivis} (bottom right). The red curve shows the best-fitting Gaussian model, with $\mu=-3.16\times10^{-3}\,\mathrm{Jy\,arcsec^{-2}}$ and $\sigma=3.22\times10^{-2}\,\mathrm{Jy\,arcsec^{-2}}$. The residual distribution is approximately Gaussian and centered close to zero.
    }
  \label{fig:hist_residual}
\end{figure}
We applied \CODE\, ({\tt Classic3D}) with $\lambda_r=1$ and a positivity constraint. 
Figure~\ref{fig:SPS_fbm_mosaic} (bottom panel) shows the same power-spectrum analysis as performed for the fBm simulation.
It is interesting to note that the power spectrum displays a power-law shape that flattens at higher $k$.
This flattening is known to arise from the contribution of the Cosmic Infrared Background (CIB) and noise components present in the image \citep{Miville-Deschenes:2010}.
In this case, the difference between linear mosaicking and joint deconvolution is less pronounced.
We find that the joint deconvolution with fusion is in very good agreement with the input mock \Herschel\, field.
This is also illustrated in Figure~\ref{fig:herschel_mosaic_ivis}, which follows the same layout as Figure~\ref{fig:fbm_linear_mosaic_ivis} and presents the results for linear mosaicking, joint deconvolution, joint deconvolution with fusion, and the corresponding residual relative to the mock \Herschel\, observation. 
Despite some structures that visually correlate with the brightest regions in the field, the residuals remain approximately Gaussian and centered close to zero, with a Gaussian fit (shown in red in Figure~\ref{fig:hist_residual}) characterized by $\mu=-3.16\times10^{-3}\,\mathrm{Jy\,arcsec^{-2}}$ and $\sigma=3.22\times10^{-2}\,\mathrm{Jy\,arcsec^{-2}}$.

\section{Application} \label{sec:application}
Having validated the method on synthetic observations of increasing complexity, we now present an application of \CODE\ (\BASE) to calibrated ASKAP visibilities from the GASKAP-HI survey.
The goals are twofold: first, to assess the performance of the method on a realistic large-scale spectral-line dataset, and second, to compare the resulting reconstructions with products currently generated using the ASKAPSoft imaging pipeline.
For this reason, and to enable a more direct comparison with the publicly available ASKAPSoft data products, we adopt a traditional feathering approach for the short-spacing correction rather than the fusion framework introduced in Section~\ref{subsec:fusion}.

\subsection{Interferometric Data}
\begin{table*}
\centering
\small
\setlength{\tabcolsep}{4pt}
\caption{Summary of the first beam (\texttt{beam00}) Measurement Sets used in this work, covering four ASKAP fields.}
\label{tab:ms_summary}
\begin{tabular}{lcccccccc}
\hline
Field & SB & Start (UTC) & End (UTC) & Duration [h] & $\nu_0$ [MHz] & $BW$ [MHz] & $B$ [kHz] & $N_{\rm vis}$ \\
\hline
MS\_M000+04A\_5 & 79298 & 2025-11-21 11:47 & 22:03 & 10.252 & 1402.00 & 4.89 & 2.315 & 869\,796 \\
MS\_M001-00A\_2 & 79331 & 2025-11-22 12:19 & 22:35 & 10.260 & 1402.00 & 4.89 & 2.315 & 863\,802 \\
MS\_M355+04A\_4 & 78680 & 2025-11-01 11:54 & 22:09 & 10.255 & 1402.00 & 4.89 & 2.315 & 862\,470 \\
MS\_M355-00A\_1 & 77022 & 2025-09-21 14:57 & 22:12 & 7.252  & 1402.00 & 4.89 & 2.315 & 632\,034 \\
\hline
\end{tabular}
\smallskip
\parbox{\textwidth}{\footnotesize
\textit{Notes.}
$\nu_0$ is the central observing frequency, $BW$ is the total processed bandwidth, and $B$ is the spectral resolution (channel width), following the notation of \citet{pingel_2022}.
$N_{\rm vis}$ corresponds to the number of calibrated visibilities for the first ASKAP beam only; a full block contains 108 beams.
}
\end{table*}

\begin{figure*}
    \centering
    \includegraphics[width=0.49\linewidth]{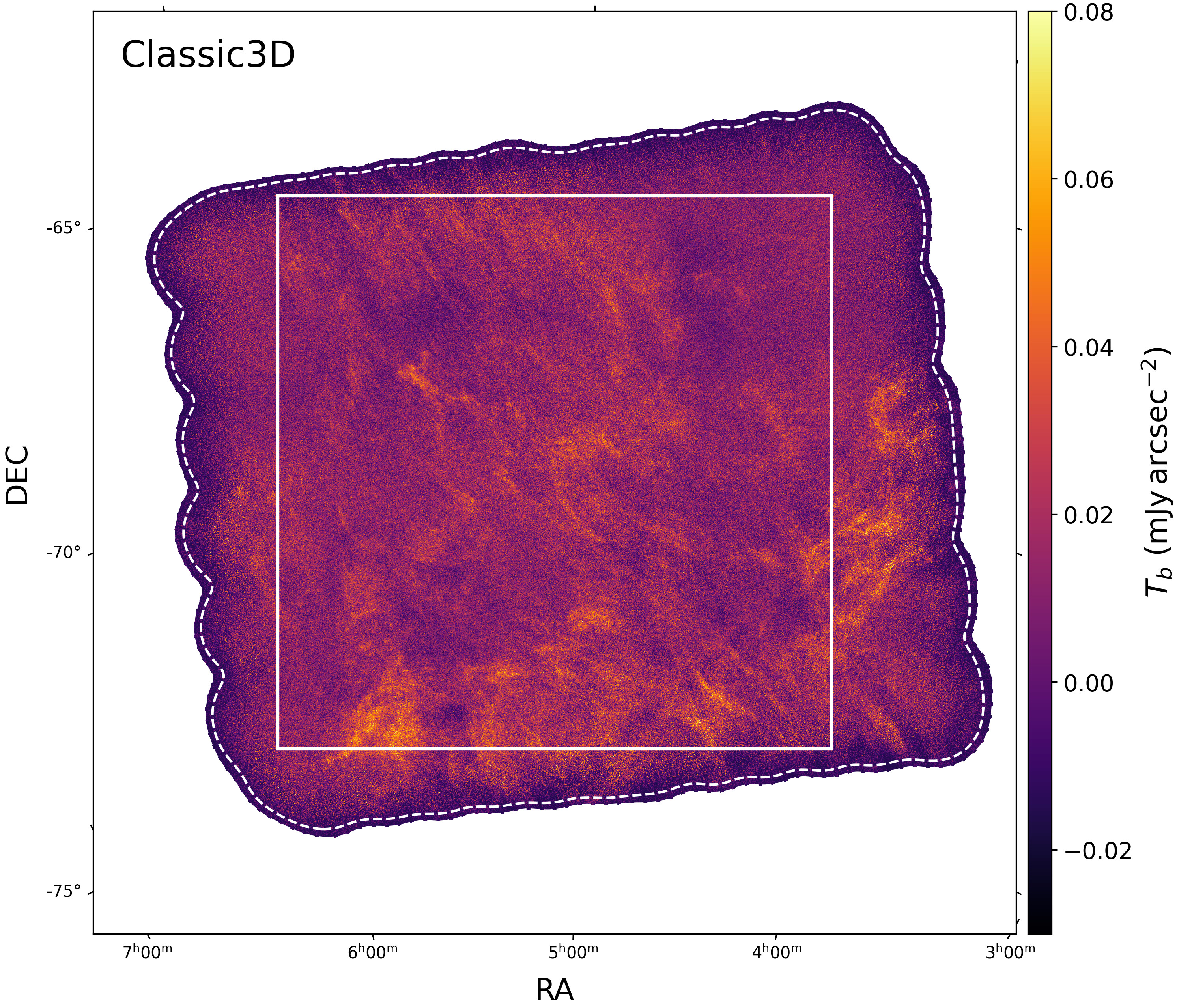}
    \includegraphics[width=0.49\linewidth]{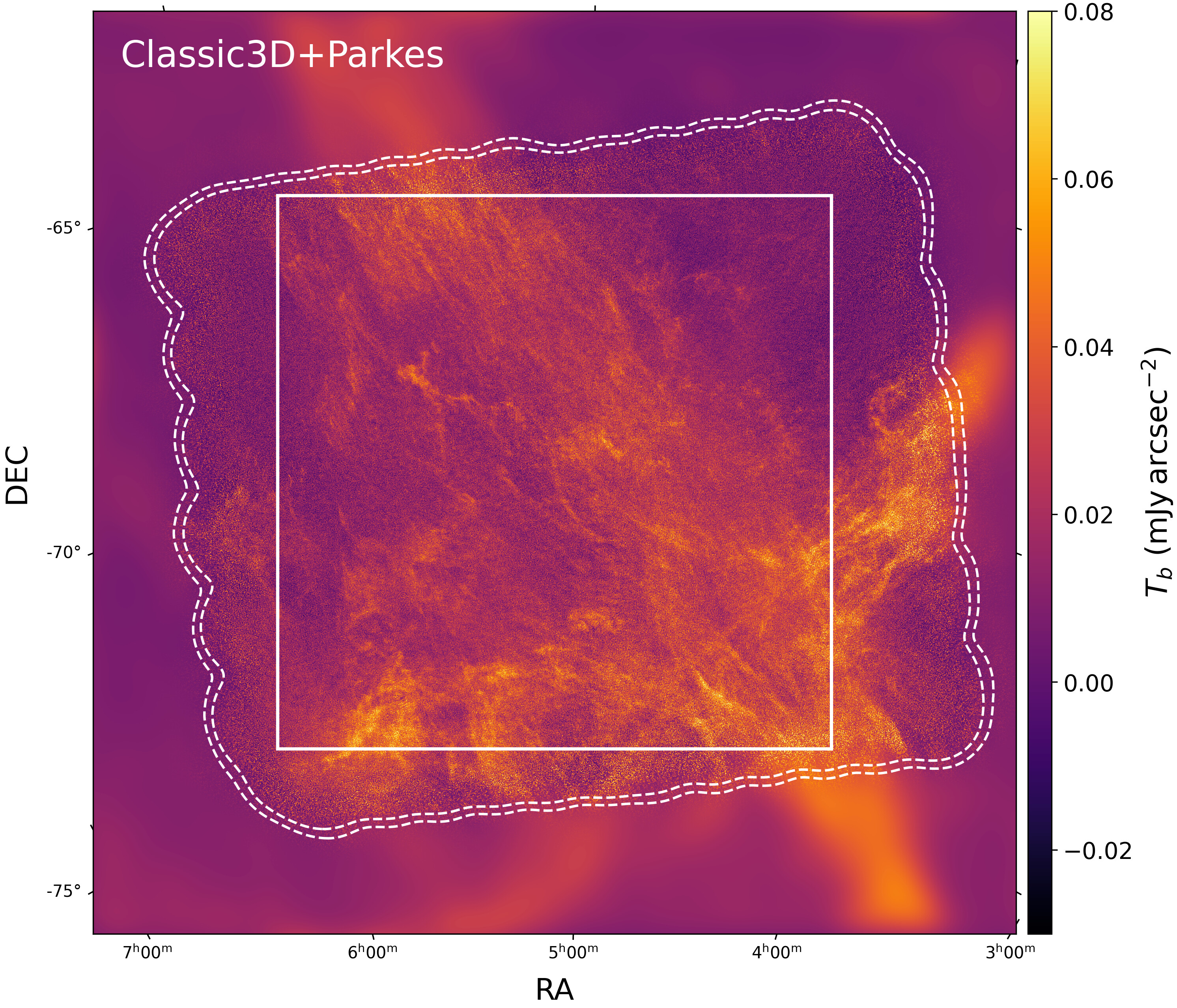}
    \includegraphics[width=0.49\linewidth]{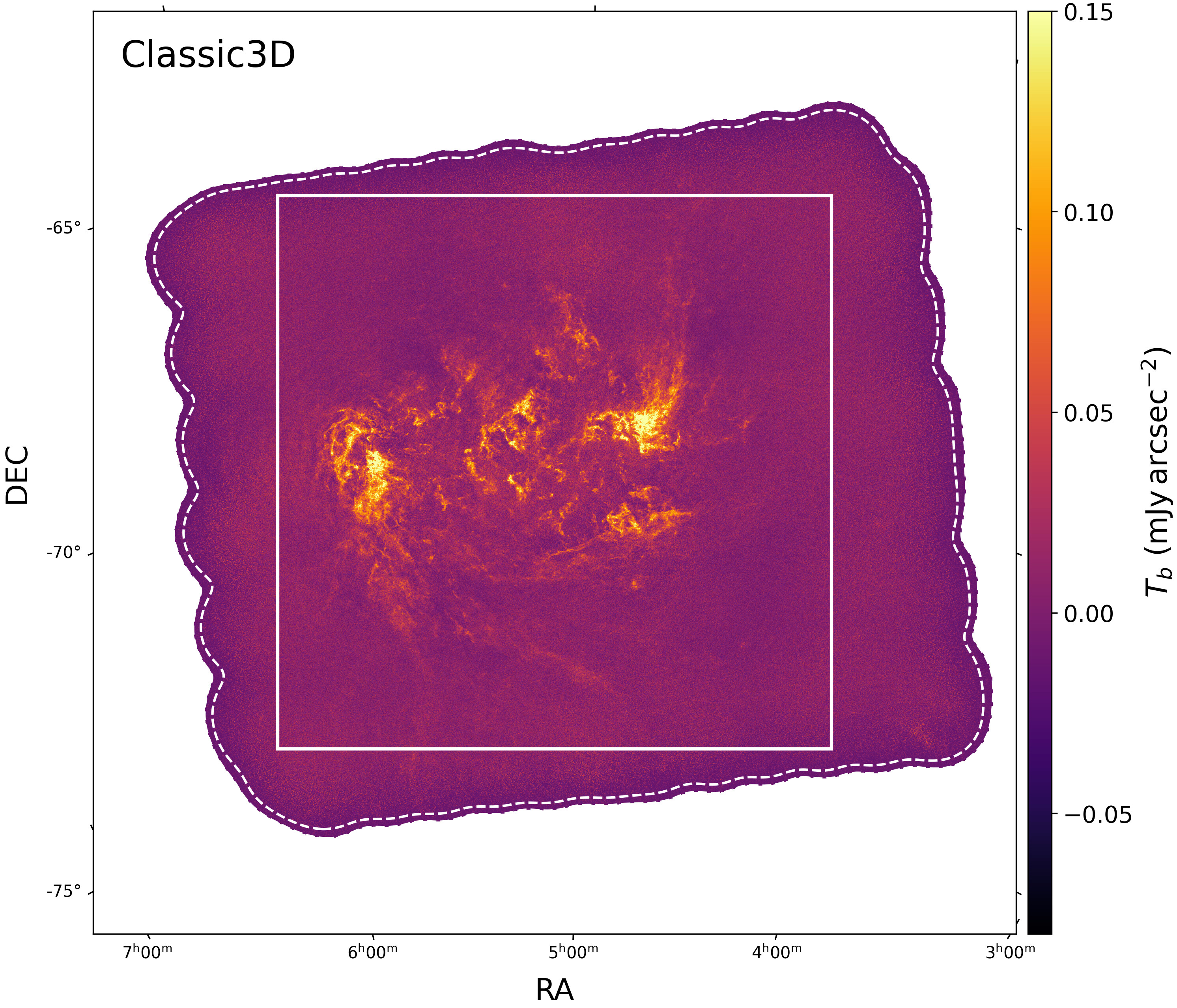}   
    \includegraphics[width=0.49\linewidth]{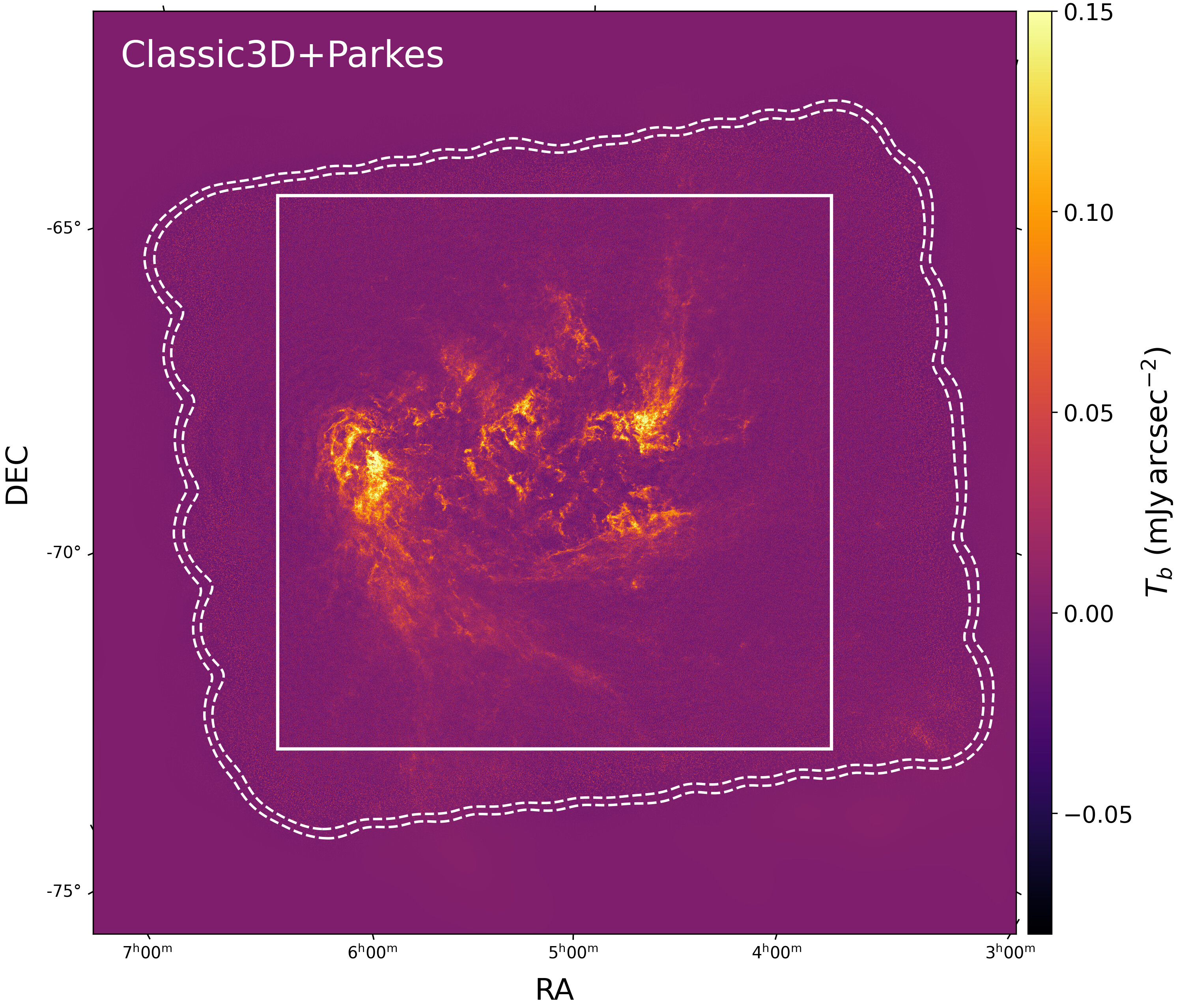}
    \includegraphics[width=0.49\linewidth]{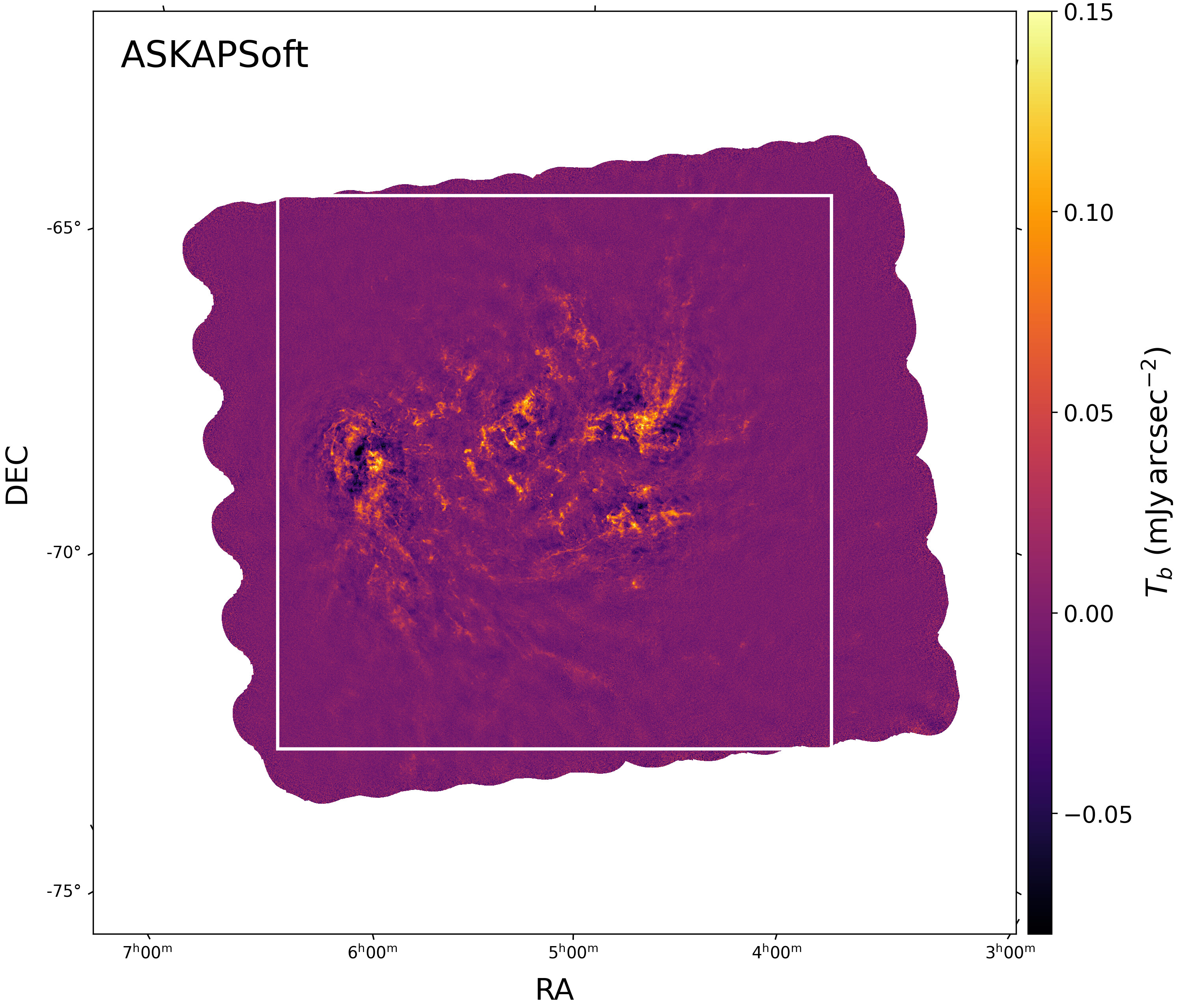}   
    \includegraphics[width=0.49\linewidth]{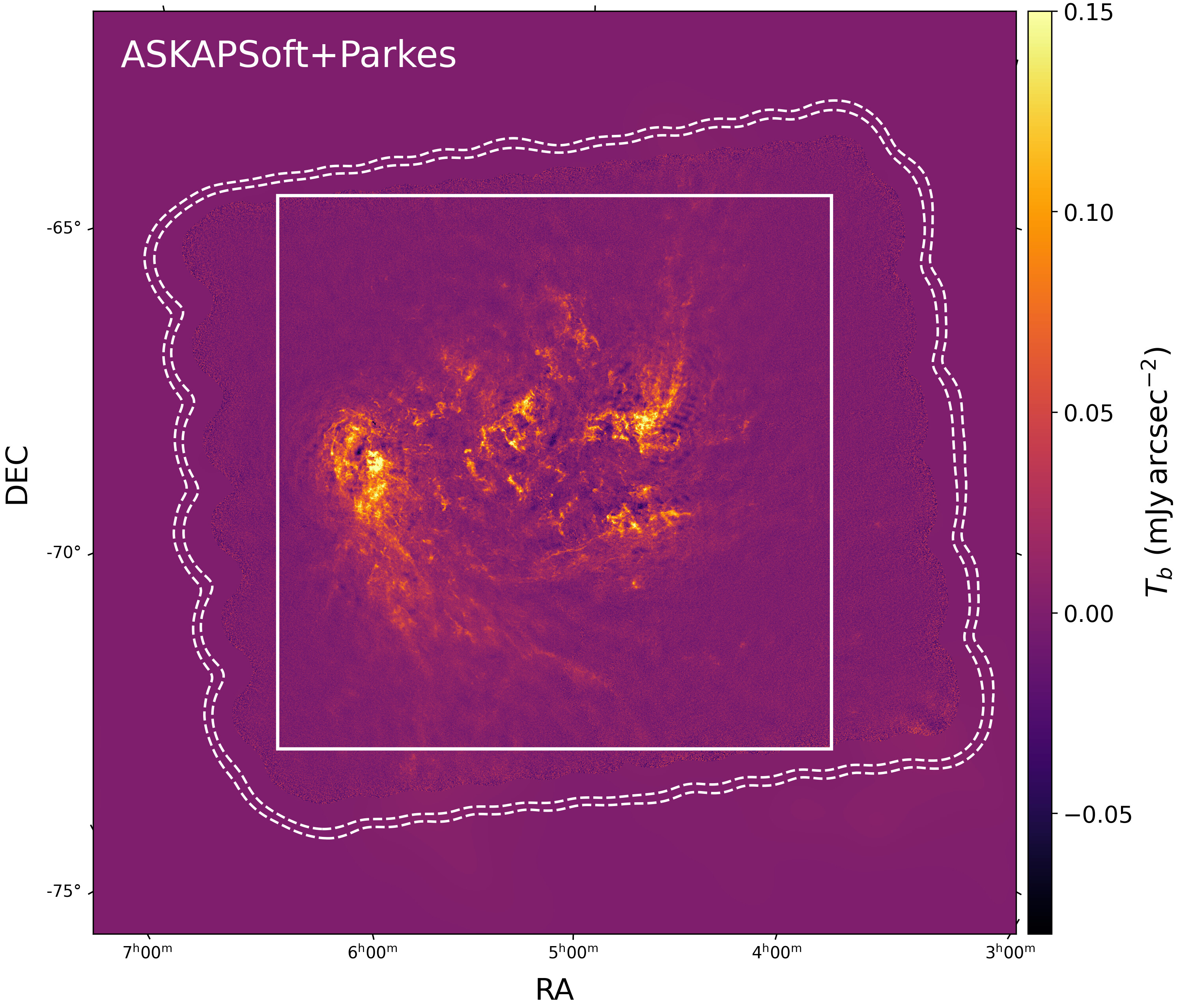}
  \caption{
    Top: ASKAP image $\hat\Ib(\rb)$ of the MW foreground toward the LMC ($10$\,h) at $v_{\rm LSR}=6.49$\,km\,s$^{-1}$ obtained with \CODE\, (\BASE) with $\lambda_r=1$ and a positivity constraint.
    The white dashed contours annotate the 0.05 and 0.1 effective primary beam levels.
    The white square outlines the region used to compute the spatial power spectra shown in Figure~\ref{fig:SPS_ASKAP}.
    The right panel shows the short-spacing corrected image obtained using traditional feathering with GASS data.
    Middle: Same but for the LMC ($10$\,h) at $v_{\rm LSR}=238.6$\,km\,s$^{-1}$.
    Bottom: Same as for the LMC velocity but using ASKAPSoft.
    }
\label{fig:output_chan_1270_vel_6.4905_2blocks_7arcsec_lambda_r_1_positivity_true_iter_20_Nw_0_ASKAP_only}
\end{figure*}

\begin{figure}
    \includegraphics[width=\linewidth]{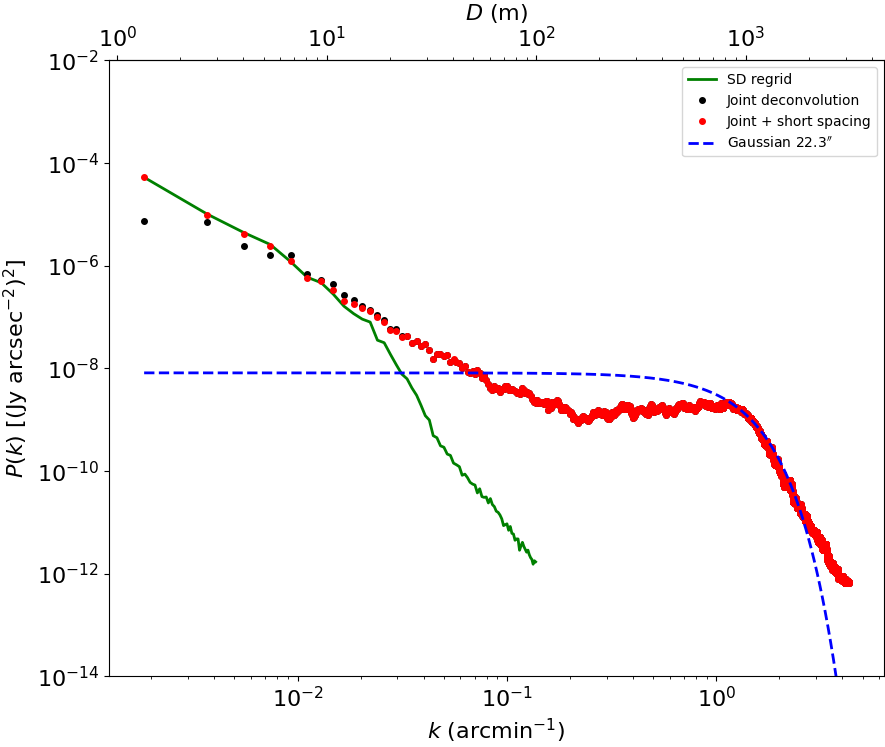}
    \includegraphics[width=\linewidth]{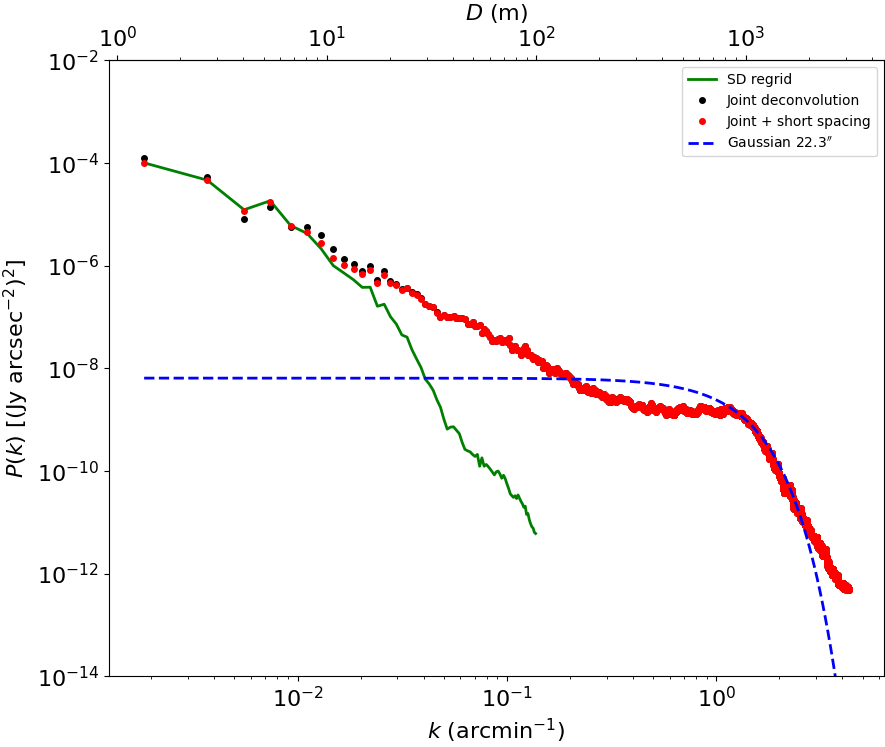}
  \caption{
    Top: Power spectra $P(k)$ (computed in the solid white box) from the ASKAP mosaic toward the MW foreground of the LMC at $v_{\rm LSR}=6.49$\,km\,s$^{-1}$.
    The black points are from the joint deconvolution performed with \CODE\, (\BASE) with $\lambda_r=1$.
    Red points show the same image but after the short-spacing correction (feathering with GASS).
    The green line shows the GASS (Parkes) observation at 16$^\prime$.
    The blue dashed line show a Gaussian beam of $\theta_{\rm FWHM}=22.3^{\prime\prime}$ (top) and $\theta_{\rm FWHM}=22.3^{\prime\prime}$ (bottom) adjusted between D=600-1600\,m.
    Bottom: Same as the top panel but for $v_{\rm LSR}=238.6$\,km\,s$^{-1}$.
    }
  \label{fig:SPS_ASKAP}
\end{figure}
\begin{figure}
    \includegraphics[width=\linewidth]{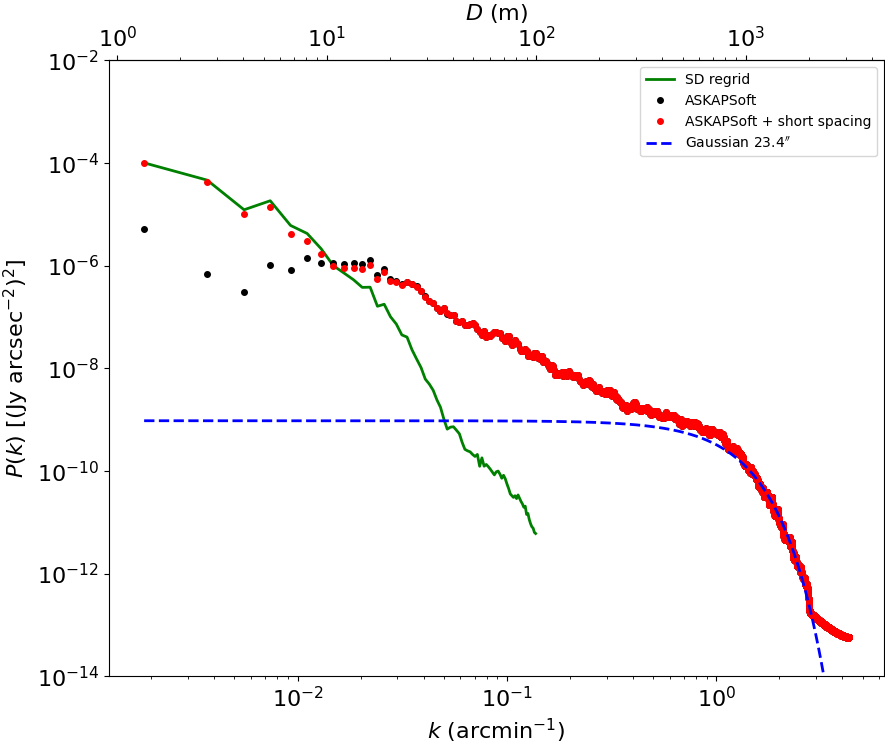}
  \caption{
    Power spectra $P(k)$ (computed in the solid white box) from the ASKAPSoft mosaic toward the LMC at $v_{\rm LSR}=238.6$\,km\,s$^{-1}$.
    The black points are from the joint deconvolution performed with ASKAPSoft (multi-scale CLEAN).
    Red points show the same image but after the short-spacing correction (feathering with GASS).
    The green line shows the GASS (Parkes) observation at 16$^\prime$.
    The blue dashed line show a Gaussian beam of $\theta_{\rm FWHM}=22.3^{\prime\prime}$ (top) and $\theta_{\rm FWHM}=23.4^{\prime\prime}$ (bottom) adjusted between D=600-1600\,m.
    }
  \label{fig:SPS_ASKAPSoft}
\end{figure}

We made use of calibrated visibilities from ASKAP that are part of the \textit{full-survey} data of the large program GASKAP-HI. 
Specifically, we used one round of four scheduling blocks\footnote{A \textit{block} refers to a single and continuous observation acquired by ASKAP.} which, once combined, covers the LMC.
Table~\ref{tab:ms_summary} gives a summary of the first beam (\texttt{beam00}) Measurement Sets.
$\nu_0$ is the central observing frequency, $BW$ is the total processed bandwidth, and $B$ is the spectral resolution (channel width), following the notation of \citet{pingel_2022}. 
$N_{\rm vis}$ corresponds to the number of calibrated visibilities for the first ASKAP beam only; a full block contains 108 beams.
Except for SB77022, which was observed for about 7.3\,h, all other blocks were observed for approximately 10.3\,h.
In total, most beams have a combined observing time of about 10\,h, and we therefore refer to the imaged mosaics according to this effective integration time.
All together, ignoring flagged baselines, the dataset for a single frequency contains approximately $272$\,M visibilities, and in this application we have produced images around two frequencies/velocities, at $v_{\rm LSR}=6.49$\,km\,s$^{-1}$ and $v_{\rm LSR}=238.6$\,km\,s$^{-1}$. These correspond the MW foreground gas, and gas in the LMC, respectively.
Continuum subtraction for these observing blocks was performed in the visibility domain using line-free channels only. Channels containing \HI emission (260–1897) were excluded from the original 2112-channel band, and the remaining 475 line-free channels at either end of the band were averaged to form a continuum image. A CLEAN model derived from this image was subtracted from the visibilities prior to spectral-line joint imaging. The final images were subsequently combined with GASS data using a traditional feathering approach following joint deconvolution with \CODE.
Finally, we use holography-derived beam images for $\Ab_{k,\nu}(\ell,m)$ provided by the ASKAP observatory. Note, however, that since we are working with spectral-line observations over a relatively narrow bandwidth, we assume $\Ab_{k,\nu}(\ell,m) \simeq \Ab_k(\ell,m)$, i.e., we neglect the small frequency dependence of the primary beam across the observed band.

\subsection{Results} \label{subsec:results}
Figure~\ref{fig:output_chan_1270_vel_6.4905_2blocks_7arcsec_lambda_r_1_positivity_true_iter_20_Nw_0_ASKAP_only} shows the results obtained with \CODE\, (\BASE) with $\lambda_r=1$ and a positivity constraint, at $v_{\rm LSR}=6.49$\,km\,s$^{-1}$ (top row) and $v_{\rm LSR}=238.6$\,km\,s$^{-1}$ (middle row), respectively. 
The pixel size of the grid was chosen to be 7$^{\prime\prime}$. 
The left panels show ASKAP only, and the right panels show ASKAP feathered with Parkes data that were interpolated at these velocities. 
On both panels, white dashed contours annotate the 0.05 and 0.1 effective primary beam levels. 
We also show in Appendix~\ref{app:mosaicking} the results obtained for linear mosaicking, as well as the difference (before feathering) between linear mosaicking and joint deconvolution. 
Interestingly, we find that the joint deconvolution solution exhibits less residual side-lobes than that of the linear mosaicking solution.

In Figure~\ref{fig:SPS_ASKAP}, we show the spatial power spectra $P(k)$ computed in the solid white box for each panel of Figure~\ref{fig:output_chan_1270_vel_6.4905_2blocks_7arcsec_lambda_r_1_positivity_true_iter_20_Nw_0_ASKAP_only} (top and middle rows only).
The solution obtained for ASKAP is shown in black, and the solution for ASKAP feathered with single-dish data from GASS is shown in red. 
We also show the $P(k)$ of the single-dish data for both velocities. 
The corresponding plot for the linear mosaicking solution is also shown in Appendix~\ref{app:mosaicking}. 
We find that for the MW foreground ASKAP mosaic is consistent with the Parkes data up to $D$ of about 10\,m. Below that, the power drops significantly. 
At the LMC velocity on the other hand, ASKAP is able to recover the correct power over most of the spatial-frequency range sampled by the interferometer. 
However, the very lowest spatial frequencies, including the total-power ($k=0$) mode that is fundamentally inaccessible to the interferometer, still require a short-spacing correction using single-dish data.
For both velocities, once combined with GASS data, these $P(k)$ both exhibits a power-law-like scale dependency. 
Quantifying their spectral indices is beyond the scope of this paper but will provide valuable insight in future analysis to understand the statistical properties of the turbulent energy cascade acting in the ISM. 

At high spatial frequencies, two effects become clearly visible: the flattening of the power spectra as the reconstruction becomes increasingly noise dominated\footnote{Unlike in Section~\ref{subsec:herschel}, there is no point-source-like emission, or continuum emission that was subtracted prior to the deconvolution.}, and the turnover imposed by the spatial regularization, which suppresses the highest spatial frequencies.
Following the methodology described in Section~\ref{sec:validation}, we fitted a Gaussian function (blue dashed line in Figure~\ref{fig:SPS_ASKAP}) with a free parameter $\theta_{\rm FWHM}$ to quantify the effective resolution of the reconstruction. This yields $\theta_{\rm FWHM}\simeq 22^{\prime\prime}$. 
We note that this corresponds to the effective resolution of the \BASE\ model solutions for $\lambda_r = 1$. Varying $\lambda_r$ would lead to a different effective resolution, while for $\lambda_r = 0$ (not shown here), the reconstruction is no longer associated with a well-defined effective beam and instead becomes constrained by the image grid itself, while also becoming increasingly dominated by correlated noise at small spatial scales.
For this specific value, the effective beam is found to be close to Nyquist sampling, as expected for a regularization strength that provides a good compromise between spatial coherence and data fidelity.
We note, however, that the flattening cannot be described simply as a white-noise plateau combined with a single Gaussian effective beam. As illustrated by the spatially varying power spectra in Figure~\ref{fig:varying_res_annuli_sps}, the reconstructed noise exhibits a scale-dependent, correlated structure, particularly away from the pointing center. The observed behavior may therefore reflect a more complex scale-dependent response, with some attenuation of spatial modes on scales larger than those indicated by the fitted effective beam.

\subsection{Comparison with ASKAPSoft (multi-scale CLEAN)}
\label{subsec:askapsoft}
We performed a direct comparison with the data products available from the CSIRO ASKAP Science Data Archive (CASDA) for the same scheduling blocks. For each block, we downloaded the continuum-subtracted image at $v_{\rm LSR}=238.6$\,km\,s$^{-1}$ produced by the {\tt ASKAPSoft} pipeline. The four images (corresponding to the four blocks tabulated in Table~\ref{tab:ms_summary}) were then re-projected and combined onto the same grid used for imaging with \CODE.
It is important to note that {\tt ASKAPSoft} encompasses the full ASKAP processing chain, including calibration and imaging. In the following, however, we use ``{\tt ASKAPSoft}'' specifically to refer to the deconvolution products generated from calibrated visibilities.
A detailed description of the multi-scale CLEAN parameters used is provided in Appendix~\ref{app:ASKAPSoft}.

The bottom row in Figure~\ref{fig:output_chan_1270_vel_6.4905_2blocks_7arcsec_lambda_r_1_positivity_true_iter_20_Nw_0_ASKAP_only} shows the {\tt ASKAPSoft} result (left), and the same image after applying the short-spacing correction used for the \CODE\ reconstruction (right).
Residual side-lobe patterns around the brightest emission regions appear more pronounced in the {\tt ASKAPSoft} image than in the \CODE\ ({\tt Classic3D}) reconstruction.
Figure~\ref{fig:SPS_ASKAPSoft} shows the corresponding $P(k)$, for comparison with the bottom panel of Figure~\ref{fig:SPS_ASKAP}. In contrast to \BASE, which recovers a substantial fraction of the low-spatial-frequency power, the {\tt ASKAPSoft} reconstruction exhibits a pronounced deficit on these scales.
Interestingly, despite relying on the coplanar-array approximation and therefore neglecting the $w$-term, the \CODE\ reconstruction exhibits reduced residual side-lobe structure compared to the corresponding {\tt ASKAPSoft} image. This suggests that, for the angular scales and observing setup considered here, the choice of deconvolution framework may have a larger impact on the final image quality than the explicit treatment of the $w$-term itself (as discussed in Section~\ref{subsec:forward}).

Finally, it is important to note that although \CODE\ appears to perform better for a single scheduling block, the imaging of \textit{full-survey} data was always intended to be performed using multiple blocks (at least three, i.e., about 30\,h) observed at different times of day. Increasing the number of visibilities would naturally reduce the discrepancy between the two methods through improved $uv$ coverage. Nevertheless, it is interesting that under more limited observing conditions (a situation that may become increasingly common in the SKA era) the \texttt{Classic3D} model appears to be more resilient.

\subsection{Comparison with previous survey}
\label{subsec:atca}
We made use of the LMC mosaic observed with the Australian Telescope Compact Array (ATCA) from \citet{Kim:1998}, that was subsequently short spacing-corrected using the Parkes telescope \citep{Kim:2001}.
Specifically, we used these data as a direct comparison with the channel imaged with \CODE\, at $v_{\rm LSR}=238.6$\,km\,s$^{-1}$. 
This comparison is particularly interesting because the ATCA+Parkes data were reconstructed using the joint-deconvolution framework introduced by \citet{Sault:1996} (see also \citealt{Kim:2003}), providing an opportunity to compare the visibility-domain approach adopted by \CODE\ with a well-established methodology that has been widely used for imaging diffuse \HI\ emission.
One caveat is that the ATCA data have a significantly lower spectral resolution than the ASKAP data used here (1.649 \kms\ for ATCA versus 0.5 \kms\ for ASKAP). We therefore interpolated the ATCA data to match the exact velocity channels of the ASKAP cube. As a result, the intensities are not expected to be strictly comparable; rather, this comparison should be regarded as a first-order diagnostic.
Note also that the ATCA+Parkes product has no continuum.

\begin{figure}
    \centering
    \includegraphics[width=\linewidth]{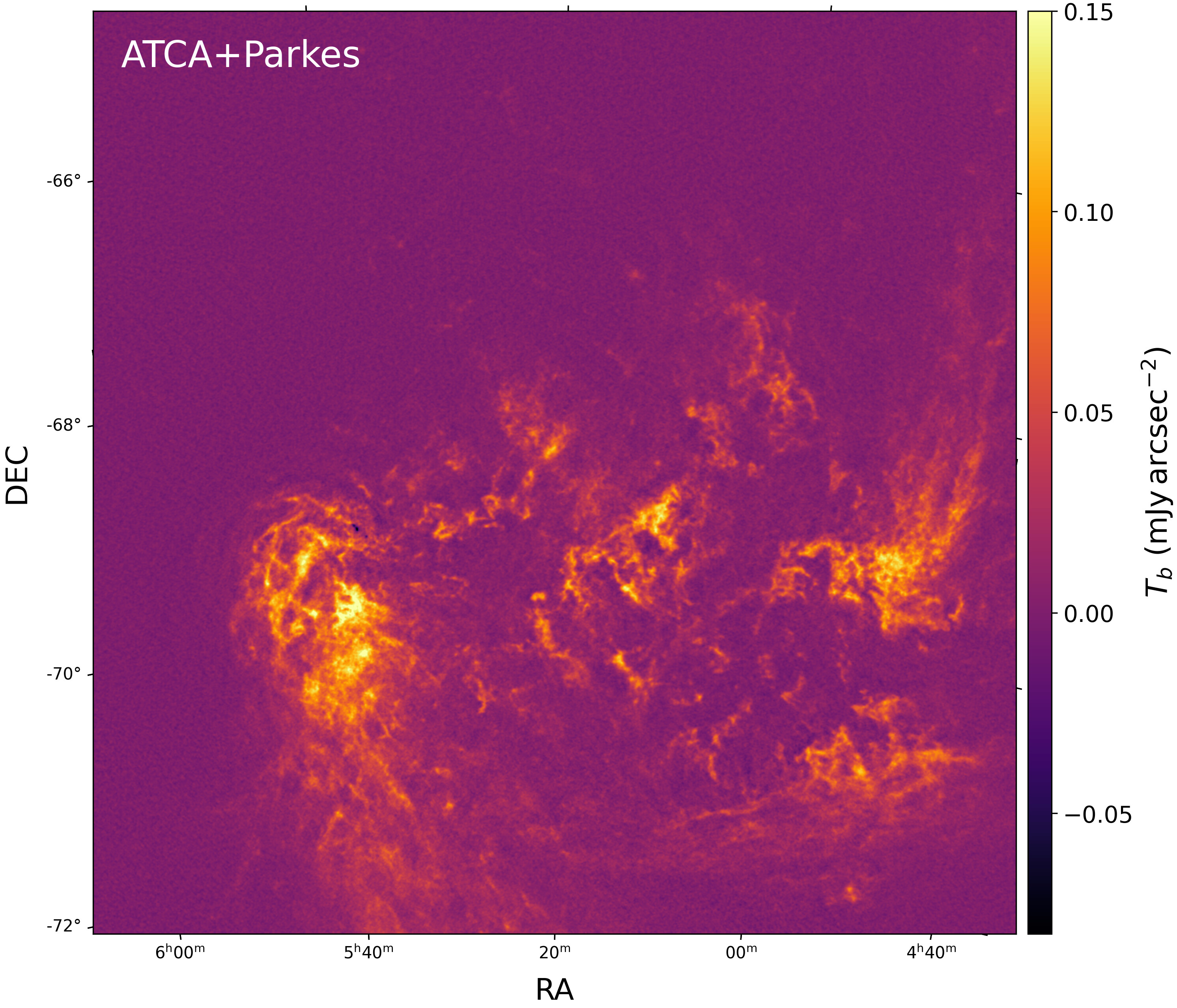}
    \includegraphics[width=\linewidth]{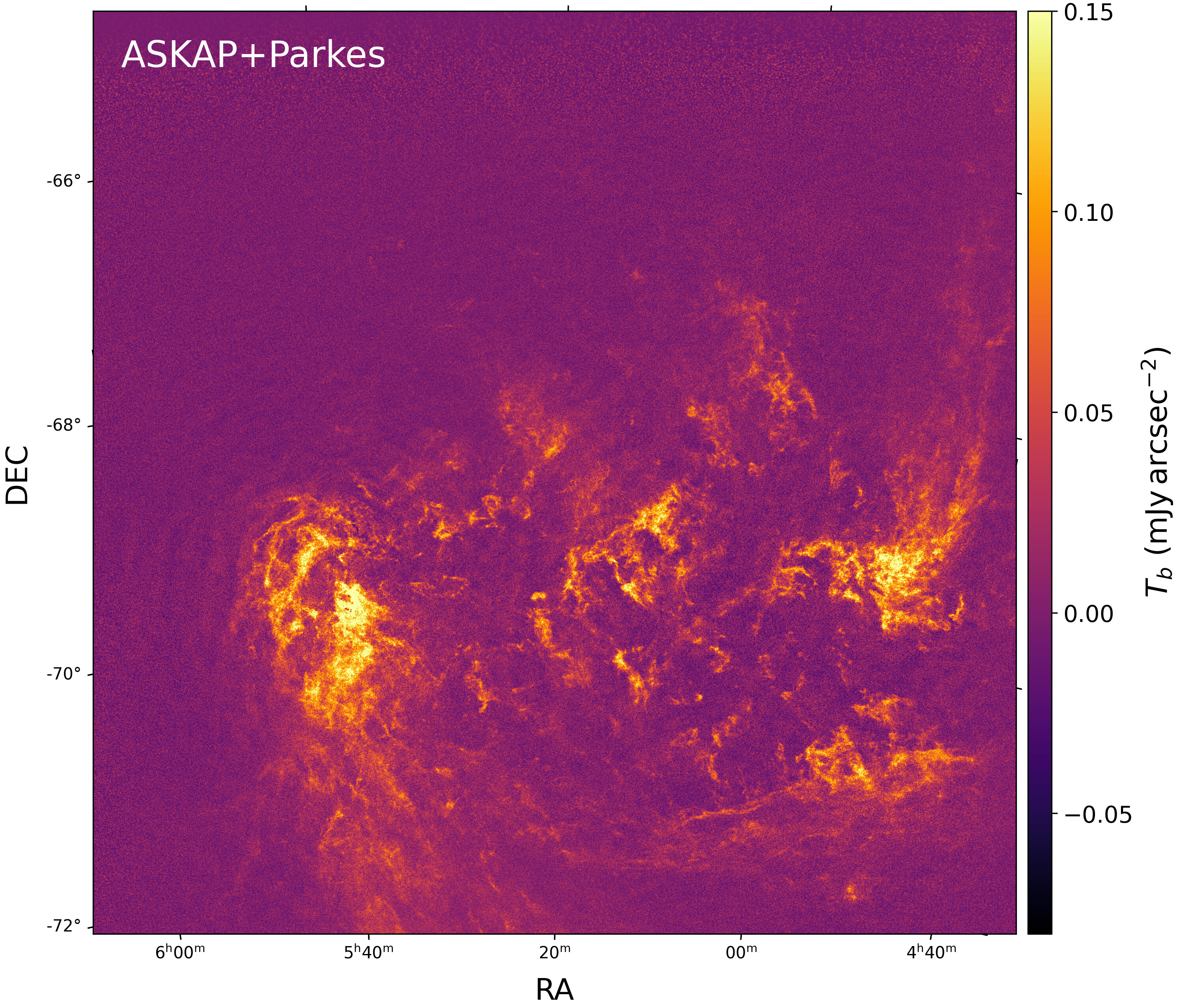}
  \caption{
    Visual comparison between ATCA and ASKAP for one channel in the LMC. 
    Top: ATCA+Parkes image of the the LMC interpolated at $v_{\rm LSR}=238.6$\,km\,s$^{-1}$.
    Bottom: ASKAP+Parkes (\CODE, \BASE) re-projected on the ATCA grid.
    } 
  \label{fig:ATCA_+Parkes_inferno}
\end{figure}

Figure~\ref{fig:ATCA_+Parkes_inferno} shows this comparison, with the ATCA+Parkes data in the top panel and the ASKAP+Parkes (\CODE, \BASE) data in the bottom panel.
Interestingly, both images show residual side-lobes in the part of the field close to the star-forming region 30~Doradus. 
These are slightly more pronounced in the ASKAP solution.
Despite this, the overall solutions look very similar, with the ASKAP result exhibiting finer structures due to its higher effective resolution ($\sim$22$^{\prime\prime}$ for ASKAP vs $\sim$1$^\prime$ for ATCA).

\subsection{Computation time}
Each image took about 12~min to optimize on a NVIDIA Tesla V100 with 32GB memory. 
The computing time being linear with the number of channels, we can predict that a full cube would require 733~GPU\,h on this architecture for a effective integration time of about 20\,h. 
The deconvolution of each channel is so far performed independently, and so this computation time can easily be distributed on multiple GPUs.
For example, using 10 GPUs would lead to a total wall time of about 73\,h (or $\sim$3~days).
These numbers are broadly comparable to current ASKAPSoft imaging times for large GASKAP-HI spectral-line cubes. 
For reference, current ASKAPSoft processing typically relies on CPU-based distributed clusters (e.g., $\sim$25 nodes with 64 CPU cores each), corresponding to roughly one spectral channel every $\sim$2.7~min for a single ASKAP observing block. 
Since the dataset considered here combines four observing blocks, this would correspond to approximately $\sim$11~min per channel, comparable to the timings obtained with \CODE. 
Direct comparisons nevertheless remain difficult because ASKAPSoft includes additional wide-field corrections such as w-stacking, which are not yet implemented in the current version of \CODE.

\section{Conclusions} \label{sec:conclusions}
This paper demonstrates that visibility-domain, non-parametric, and GPU-accelerated joint deconvolution is computationally feasible for modern wide-field spectral-line datasets and provides a practical framework for the reconstruction of large-scale diffuse \HI\ emission.
The \CODE\ framework, and its base implementation \BASE, enable direct inversion of calibrated visibilities while naturally incorporating spatial regularization and, when available, single-dish information through a unified optimization scheme.
The main findings are as follows:
\begin{itemize}
    \item We show that \CODE\ accurately recovers point-source flux in the absence of regularization ($\lambda_r=0$).
    
    \item We demonstrate that the Laplacian regularization controls the trade-off between noise suppression and effective resolution, defining an implicit \textit{effective beam} that can be characterized through power-spectrum analysis.
    
    \item We find that the noise response of the reconstruction is spatially modulated by the primary beam, leading to increased noise amplitude and larger-scale fluctuations away from the pointing center.
    
    \item Using synthetic multiscale fBm simulations, we show that joint deconvolution recovers more large-scale power than linear mosaicking of independently deconvolved pointings, consistent with the results of \citet{Cornwell:1988}. For the more realistic \texttt{Herschel}-based simulation, the difference between the two approaches is less pronounced.
    
    \item We demonstrate that incorporating single-dish information through fusion allows the reconstruction to recover the correct spatial power spectrum over a broad range of scales, including the lowest spatial frequencies.
    
    \item Applied to GASKAP-HI data, \CODE\ produces high-fidelity mosaics with reduced residual side-lobe structure and realistic multiscale morphology for both Galactic foreground emission and LMC emission.

    \item Comparison with publicly available {\tt ASKAPSoft} products (using multi-scale CLEAN) indicates broadly consistent large-scale morphology, while the \CODE\ reconstruction shows less prominent residual side-lobe patterns and recovers substantially more power at low spatial frequencies.

    \item A comparison with previous ATCA+Parkes observations of the LMC shows broad agreement in the recovered large-scale morphology, demonstrating consistency between the visibility-domain reconstruction implemented in \CODE\ and established joint-deconvolution approaches for diffuse \HI\ imaging.
    
    \item We find that the reconstructed power spectra ``once short-spacing corrections are applied'' exhibit power-law-like behavior characteristic of interstellar turbulence.
    
    \item The computational cost scales linearly with the number of spectral channels and can be efficiently distributed across GPUs, making the method suitable for large surveys such as GASKAP-HI and future SKA datasets.
\end{itemize}

Taken together, these results highlight several advantages of visibility-domain joint deconvolution for wide-field spectral-line imaging. By fitting a single sky model directly to all calibrated visibilities, information from overlapping pointings contributes simultaneously to the reconstruction, avoiding the need to combine independently deconvolved images through linear mosaicking. Likewise, the fusion formalism allows short-spacing information to constrain the solution throughout the optimization process rather than being introduced through a post-processing combination. 
More generally, the framework avoids the need for a parametric component-based representation of the sky and instead reconstructs a non-parametric image directly from the visibilities while allowing positivity constraints and spatial regularization to be treated independently. These characteristics make visibility-domain inverse-problem approaches particularly attractive for the reconstruction of complex multiscale diffuse emission.

\section{Future prospects}
\label{sec:prospects}
While the \BASE\ model presented here relies on a relatively simple regularized least-squares formulation, the modular architecture of \CODE\ was designed to facilitate the development and comparison of more advanced visibility-domain imaging strategies.
Rather than targeting a specific deconvolution methodology, the framework provides a flexible environment in which alternative forward models, likelihood functions, and regularization schemes can be implemented and evaluated on large spectral-line mosaics.

Future developments could include improved treatment of wide-field effects (e.g., explicit handling of the $w$-term and more general direction-dependent effects), more robust likelihood formulations to better handle residual interference and calibration imperfections, and additional regularization schemes such as sparsity-promoting, multiscale, or scattering-transform priors.
The framework also provides a natural setting in which physically motivated priors can be incorporated directly into the imaging process.
Beyond the single-channel applications presented here, \CODE\ was designed to support intrinsically three-dimensional optimization problems and therefore provides a natural framework for future developments in multi-frequency synthesis, low-rank spectral modeling, and full-Stokes imaging.
More generally, these developments will enable the exploration of increasingly sophisticated inverse-problem formulations while retaining the ability to perform joint visibility-domain deconvolution across large interferometric mosaics.

\section*{Data Availability}
\CODE\, source code and a \textit{Read the Docs} documentation are available in free access via the following web pages: \url{https://github.com/antoinemarchal/ivis} and \url{https://ivis-dev.readthedocs.io}. 
While \CODE\, is open source, the GASKAP-HI visibility data and imaging products are proprietary to the ASKAP Observatory and the GASKAP-HI collaboration. 

\begin{acknowledgements}
The authors thanks the Software team at CSIRO; Tim Galvin, Stephen Ord, Daniel Mitchell, Wasim Raja, Max Voronkov, Mark Wieringa, and Matthew Whiting, as well as and Alec Thomson for insightful discussions during JohnFest (ATNF, 2025). 
AM also acknowledges the ATNF Radio School (Narrabri, 2023) and the NRAO Workshop on Synthesis Imaging for Radio Interferometry (SSMID, NRAO, Charlottesville, 2024) where the \texttt{MPol} framework was introduced to AM.
We thank Benjamin Godard for providing access to the \textit{totoro} storage infrastructure.
We thank Cameron Van Eck for extensive and fruitful discussions on the theory of radio-interferometry.
Fruitful discussions with Eric Koch are also gratefully acknowledged.
AM thanks Karlie A. Noon and Enrico M. Di Teodoro for providing us with a MeerKAT dataset used in the online tutorial.
GASKAP-HI is partially funded by the Australian Government through an Australian Research Council Australian Laureate Fellowship (project number FL210100039 awarded to NM-G).
We thank the anonymous referee whose comments and suggestions have improved this manuscript.
\end{acknowledgements}

\bibliographystyle{aa}
\bibliography{aa61540-26.bib}

\begin{appendix}

\section{Linear mosaicking of ASKAP pointings} \label{app:mosaicking}
\begin{figure}
    \includegraphics[width=\linewidth]{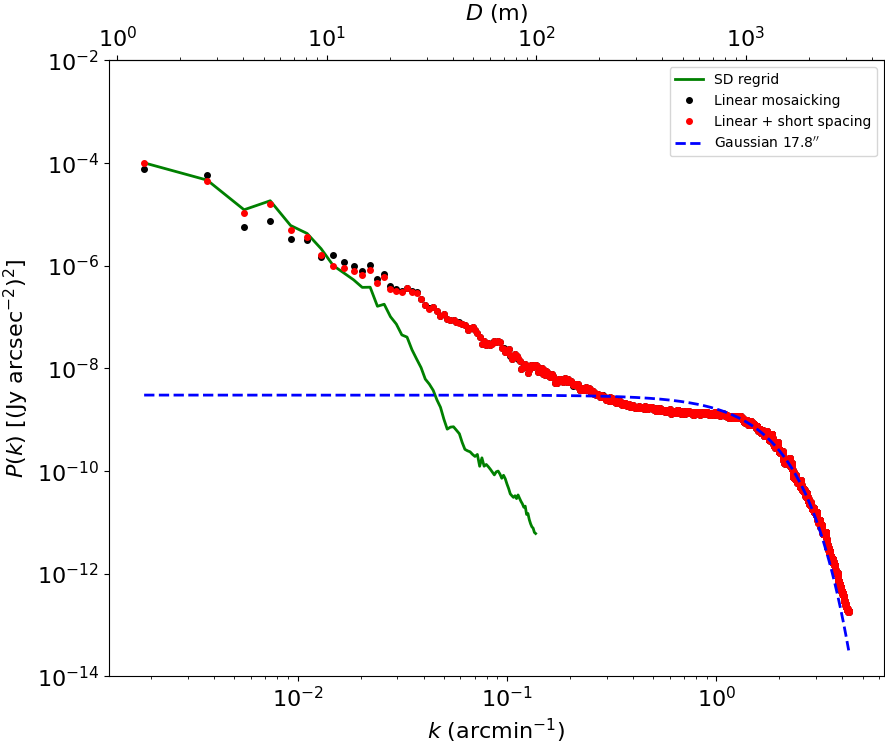}
  \caption{
    Power spectra $P(k)$ from an ASKAP mosaic (8 blocks; 4 field pointings) of observations described in Table~\ref{tab:ms_summary} toward the LMC, but for independently deconvolved pointings assembled with linear mosaicking.
    The black points show the ASKAP only solution.
    Red point show the same image but after the short-spacing correction (feathering with GASS).
    The green line shows the GASS (Parkes) observation at 16$^\prime$.
    The blue dashed line show a Gaussian beam of $\theta_{\rm FWHM}=17.8^{\prime\prime}$ adjusted between D=600-1600\,m.
    }
  \label{fig:SPS_ASKAP_linear}
\end{figure}
\begin{figure}
    \centering
    \includegraphics[width=0.96\linewidth]{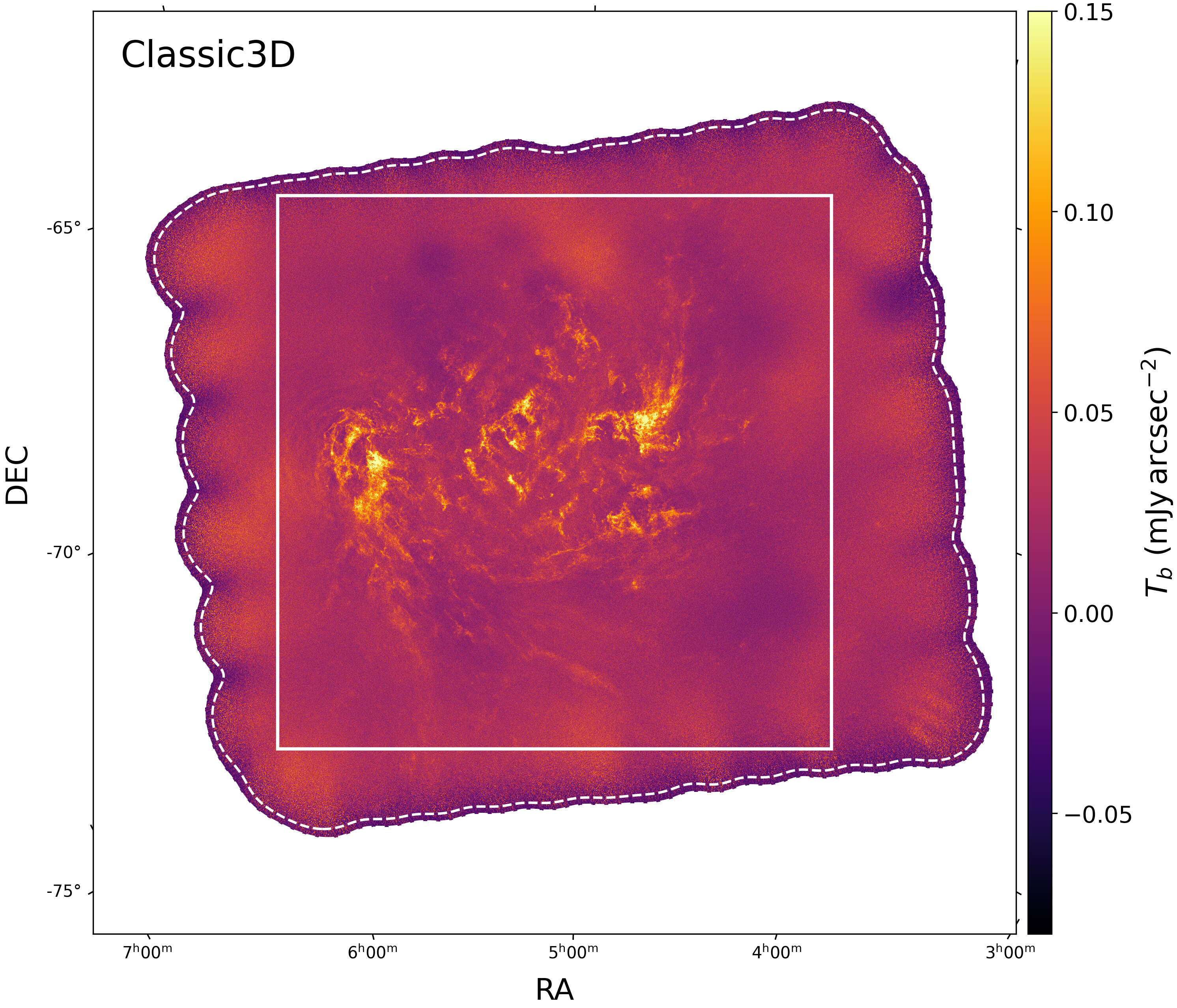}
    \includegraphics[width=0.96\linewidth]{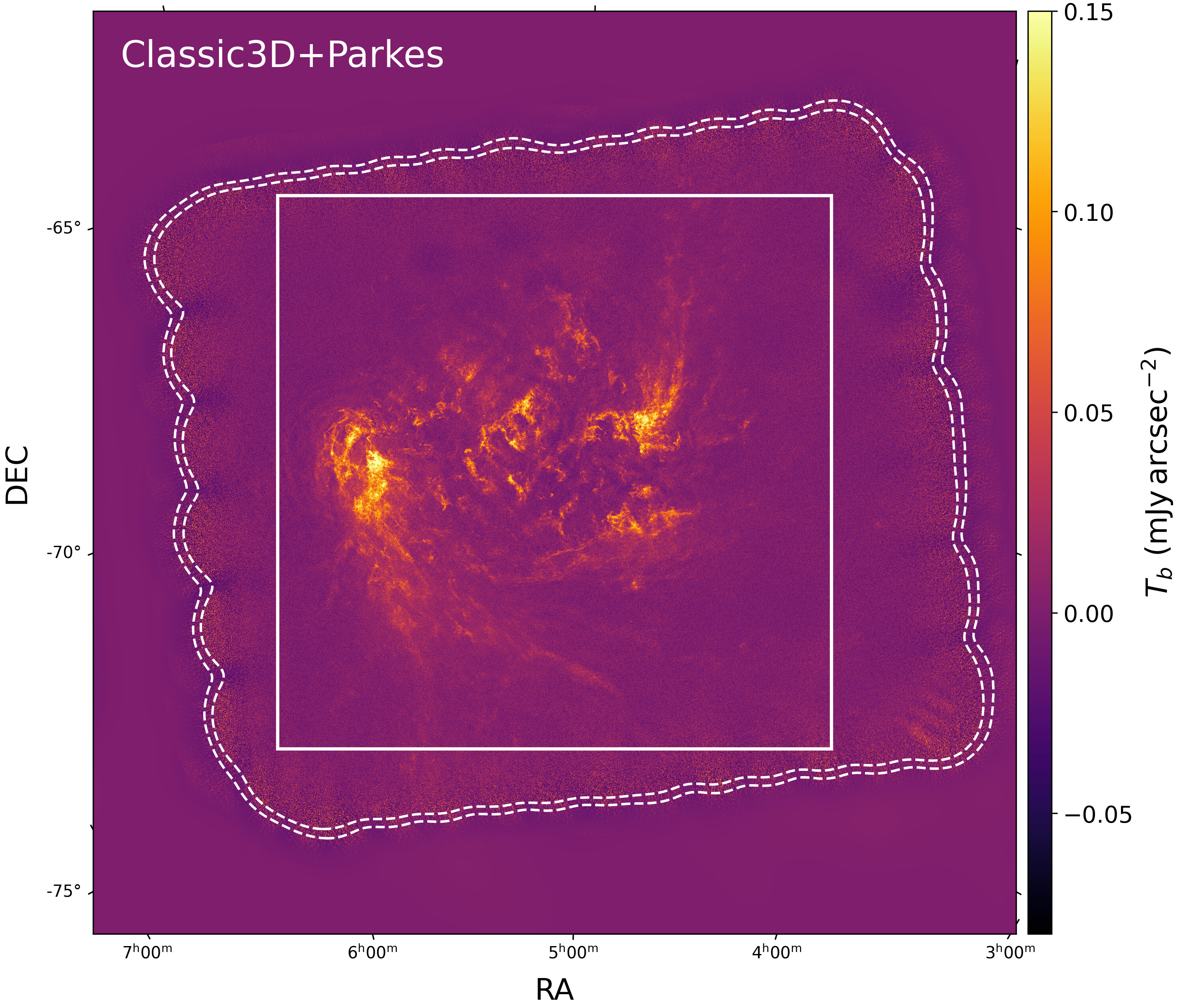}
    \includegraphics[width=0.96\linewidth]{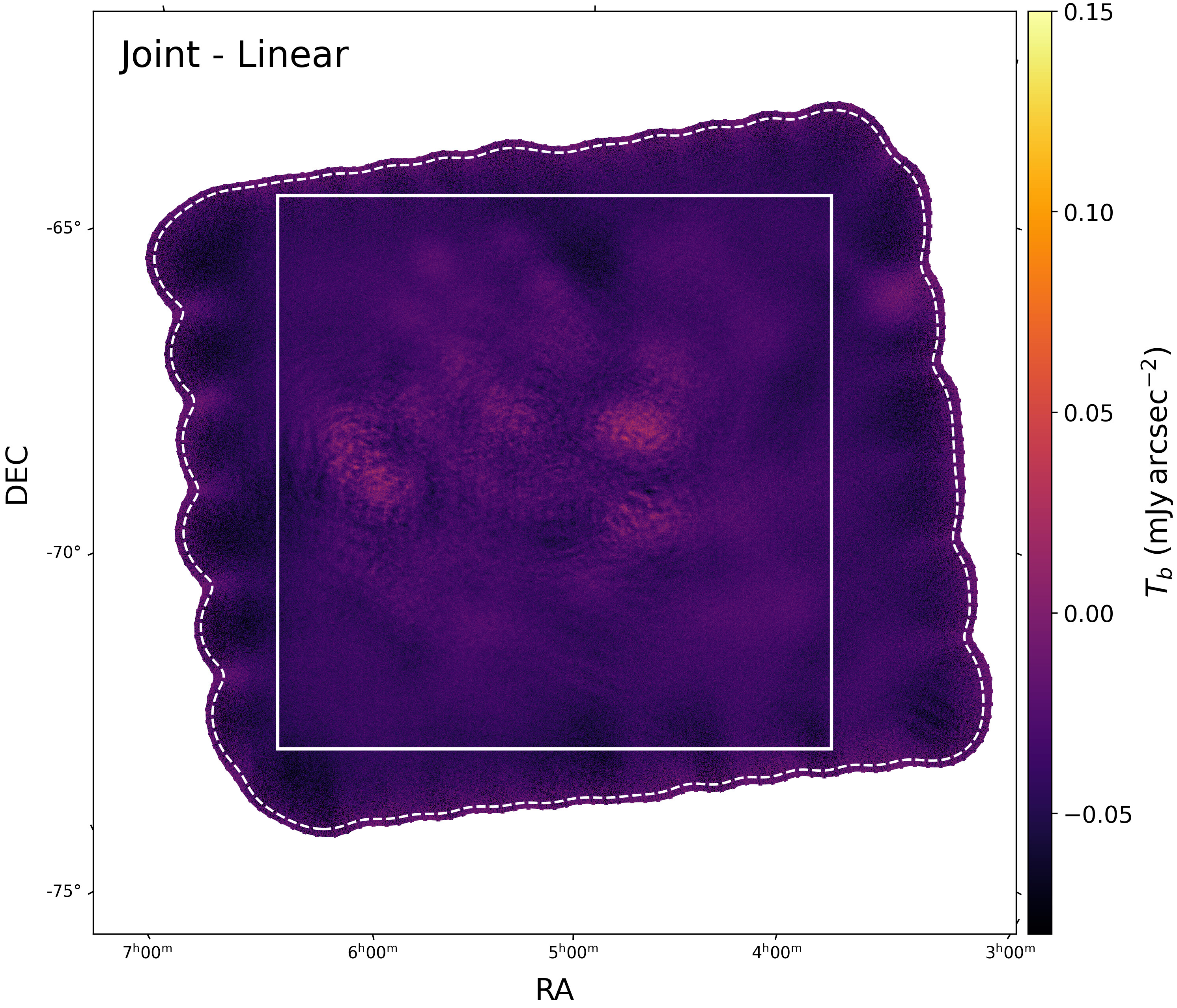}
  \caption{
    Top: ASKAP image $\hat\Ib(\rb)$ of the LMC ($10$\,h) at $v_{\rm LSR}=238.6$\,km\,s$^{-1}$ obtained with \CODE\, (\BASE) with $\lambda_r=1$ and a positivity constraint on each individual pointings that were then combined with linear mosaicking. 
    Middle: A short spacing correction was applied using traditional feathering with GASS data. 
    White dashed contours are as in Figure~\ref{fig:output_chan_1270_vel_6.4905_2blocks_7arcsec_lambda_r_1_positivity_true_iter_20_Nw_0_ASKAP_only}.
    The white square outlines the region used to compute the spatial power spectra shown in Figure~\ref{fig:SPS_ASKAP_linear}.
    Bottom: Difference between the ASKAP only solution obtained with joint deconvolution and that of this linear mosaicking. 
    }
\label{fig:output_chan_795_2blocks_7arcsec_lambda_r_1_positivity_true_iter_20_LINEAR_short_spacing}
\end{figure}
Figure~\ref{fig:SPS_ASKAP_linear} shows the power spectra of the ASKAP mosaic at LMC velocities, in the case where each pointing was deconvolved independently and then combined with linear mosaicking.
The top panel in Figure~\ref{fig:output_chan_795_2blocks_7arcsec_lambda_r_1_positivity_true_iter_20_LINEAR_short_spacing} shows the result obtained with \CODE\, (\BASE) with $\lambda_r=1$ and a positivity constraint, the middle panel shows the same image but feathered with GASS data, and the bottom panel shows the difference between the ASKAP only solution obtained with joint deconvolution and that of this linear mosaicking shown in the top panel.

\section{Short-spacing correction}
Let $I_{\mathrm{int}}(\boldsymbol{r})$ be the interferometric image and $I_{\mathrm{SD}}(\boldsymbol{r})$ the corresponding single-dish map, regridded onto the interferometric image plane.
To ensure consistent units, maps expressed in brightness temperature are converted to surface brightness through the Rayleigh--Jeans relation,
\begin{equation}
    I_\nu = \frac{2 k_{\mathrm B}\nu^2}{c^2}\,T_{\mathrm B},
\end{equation}
and are expressed in $\mathrm{Jy\,arcsec^{-2}}$.
The short-spacing correction is then performed in Fourier space. Defining
\begin{equation}
    \widetilde{I}_{\mathrm{SD}}(\boldsymbol{k})
    = \mathcal{F}\!\left[I_{\mathrm{SD}}(\boldsymbol{r})\right],
    \qquad
    \widetilde{I}_{\mathrm{int}}(\boldsymbol{k})
    = \mathcal{F}\!\left[I_{\mathrm{int}}(\boldsymbol{r})\right],
\end{equation}
the single-dish contribution is weighted by a Gaussian transfer function,
\begin{equation}
    B(\boldsymbol{k})
    = \exp\!\left[-2\pi^2 \sigma^2 |\boldsymbol{k}|^2\right],
    \qquad
    \sigma = \frac{\theta}{2\sqrt{2\ln 2}},
\end{equation}
where $\theta$ is a user-defined characteristic scale that controls the transition between single-dish and interferometric data in Fourier space.
In the implementation used here, we adopt $\theta = 0.5^\circ = 30'$. 
This value is chosen empirically based on the power-spectrum analysis (see Section~\ref{subsec:results}), and is therefore not necessarily equal to the nominal single-dish beam (e.g. 16$^\prime$).
The combined Fourier plane is written as
\begin{equation}
    \widetilde{I}_{\mathrm{comb}}(\boldsymbol{k})
    =
    B(\boldsymbol{k})\,\widetilde{I}_{\mathrm{SD}}(\boldsymbol{k})
    +
    \left[1 - B(\boldsymbol{k})\right]\widetilde{I}_{\mathrm{int}}(\boldsymbol{k}),
\end{equation}
so that the single-dish data supply the large-scale emission, while the interferometric image provides the small-scale structure.
The final short-spacing-corrected image is obtained through the inverse Fourier transform,
\begin{equation}
    I_{\mathrm{comb}}(\boldsymbol{r})
    =
    \mathcal{F}^{-1}\!\left[\widetilde{I}_{\mathrm{comb}}(\boldsymbol{k})\right].
\end{equation}

\section{Multi-scale CLEAN within ASKAPSoft} \label{app:ASKAPSoft}
Images downloaded from CASDA were produced using \texttt{ASKAPsoft} with a spatial resolution of ~30$^{\prime\prime}$ and pixel size of 6$^{\prime\prime}$ over a 4500 x 4500 pixel field. 
Gridding utilizes the A-Project W-Stack algorithm, which accounts for the wide-field w-term and primary beam effects across ASKAP's 36-feed phased array, with 9 w-planes and a Robustness parameter of 0.5 (Briggs weighting, intermediate between natural and uniform). 
A Gaussian taper of 20$^{\prime\prime}$ is applied as a preconditioner to downweight long baselines and improve sensitivity to extended emission, and a maximum $uv$ distance of 2000\,m is imposed to further suppress small-scale structure. 
Deconvolution uses the multi-scale CLEAN algorithm, which simultaneously fits emission at four angular scales defined by delta-functions and Gaussians of width 0, 6, 15, and 38 pixels — roughly corresponding to point sources and structures of $\sim$0.6$^\prime$, 1.5$^\prime$, and 3.8$^\prime$ at the adopted cell size.

In multi-scale CLEAN, rather than subtracting a single point-source component per iteration as in classical CLEAN, the algorithm identifies the scale at which the residual image is best described at each step and subtracts a scaled basis function at that size, which is more effective at recovering smooth, diffuse \HI emission without the ``bowl'' artifacts that plague single-scale CLEAN. 
The loop gain is set to 0.2, meaning 20\% of the peak component is subtracted per minor-cycle iteration, with a maximum of 3000 iterations and 10 major cycles. Cleaning proceeds down to a minor-cycle threshold of 35\%, 5\,$\sigma$, or 0.5\,$\sigma$ (whichever is reached first), with a major-cycle threshold of 10\,mJy. The restored image uses a beam fitted to the point spread function at the mid-frequency reference channel.

\end{appendix}

\end{document}